\documentclass[12pt,longbibliography]{article}

\usepackage{amssymb,amsmath}
\usepackage{graphicx}
\usepackage{mathtools}
\usepackage{xcolor}
\usepackage{diagbox,multirow,booktabs}
\usepackage[colorlinks=true,linkcolor=black,citecolor=blue]{hyperref}

\usepackage{tikz}
\usetikzlibrary{cd}
\usetikzlibrary{shapes,fit,arrows,positioning,quotes}

\renewcommand{\baselinestretch}{1.2}
\def\ba#1\ea{\begin{align}#1\end{align}}
\def\be#1\ee{\begin{equation}#1\end{equation}}
\def\bg#1\eg{\begin{gather}#1\end{gather}}
\def\bea #1\eea{\begin{equation} \begin{aligned}#1\end{aligned} \end{equation}}
\def\bpm{\begin{pmatrix}}
\def\epm{\end{pmatrix}}

\newcommand{\eg}{\textit{e.g.}}

\newcommand{\nn}{\nonumber}

\DeclareMathOperator{\tr}{tr}

\newcommand{\ep}{\epsilon}

\newcommand{\cC}{\mathcal{C}}

\newcommand{\cN}{\mathcal{N}}
\newcommand{\cO}{\mathcal{O}}

\newcommand{\cQ}{\mathcal{Q}}

\newcommand{\cT}{\mathcal{T}}

\newcommand{\bZ}{\mathbb{Z}}

\newcommand{\su}[1]{{\rm SU}(#1)}
\newcommand{\uu}[1]{{\rm U}(#1)}
\newcommand{\so}[1]{{\rm SO}(#1)}
\newcommand{\Sp}[1]{{\rm Sp}(#1)}

\renewcommand{\arraystretch}{1}

\allowdisplaybreaks

\begin{document}

\makeatletter \@addtoreset{equation}{section} \makeatother
\renewcommand{\theequation}{\thesection.\arabic{equation}}
\renewcommand{\thefootnote}{\alph{footnote}}

\begin{titlepage}

\begin{center}

\vspace*{2cm}

{\Large\bf Black hole cohomologies in $\mathcal{N}=4$ SYM}

\vspace{1cm}

\renewcommand{\thefootnote}{\alph{footnote}}

{\large Seunggyu Kim$^{1,2}$, Jehyun Lee$^{2}$, Siyul Lee$^{3}$, and Jaewon Song$^{2}$}

\vspace{0.3cm}

\vspace{.1cm}
{$^1$ Beijing Institute of Mathematical Sciences and Applications (BIMSA) \\
Huaibei Town, Huairou District, Beijing 101408, China}\par

\vspace{.1cm}
{$^{2}$ Department of Physics, Korea Advanced Institute of Science and Technology\\
291 Daehak-ro, Yuseong-gu, Daejeon 34141, Republic of Korea}\par

\vspace{.1cm}
{$^{3}$ Instituut voor Theoretische Fysica, KU Leuven, Celestijnenlaan 200D, 3001 Leuven, Belgium}\par

\vspace{0.7cm}

E-mails: {\tt kikksk456@gmail.com, 2wpgus@kaist.ac.kr, siyul.lee@kuleuven.be, jaewon.song@kaist.ac.kr}

\vspace{0.8cm}
\textbf{Abstract}
\end{center}
We systematically investigate BPS cohomologies of $\mathcal{N}=4$ super-Yang-Mills (SYM) theory with gauge group $\su{N}$ with $N=2, 3, 4$, focusing primarily on the BMN sector. 
We find non-Coulomb graviton cohomologies for $N>2$.
For $N>2$, the lightest black hole cohomology in the BMN sector always carries non-trivial $R$-charge and corresponds to the brane-fused type. Therefore, the `core' black holes without hair are `heavier'.
We also find that a family of brane-fused cohomologies in $N=3$ is not lifted at the first quantum order.
At a fixed charge sector, gravitons carry definite parity under charge conjugation. 
Up to the levels we have tested, every black hole cohomology either has opposite parity to that of gravitons or is lifted by quantum corrections.
We conjecture that black hole cohomologies always carry parity opposite to gravitons in the same charge sector.

\end{titlepage}

\renewcommand{\thefootnote}{\arabic{footnote}}

\setcounter{footnote}{0}

\renewcommand{\baselinestretch}{1}

\tableofcontents

\renewcommand{\baselinestretch}{1.2}

\pagebreak

\section{Introduction}

A black hole carries an enormous amount of entropy proportional to its area \cite{Bekenstein:1973ur, Hawking:1975vcx}. The microscopic states that account for such entropy have been studied in numerous settings, starting from a seminal work by Strominger and Vafa \cite{Strominger:1996sh}. Even though it is now well understood how these large degeneracies contributing to the black hole entropy arise from various string-theoretic setups, it still remains an open question to precisely characterize the quantum states that comprise a black hole. 

Recently, this question has started to be addressed in the context of the AdS/CFT correspondence. In AdS/CFT, any quantum state in the bulk corresponds to an operator in the boundary CFT.  
BPS states or operators in superconformal field theories (SCFTs) provide an ideal laboratory for investigating microstates. Instead of looking at individual operators, we focus on BPS cohomology \cite{Grant:2008sk, Chang:2013fba}. The BPS cohomology is defined as the operators that are closed under some supercharge $Q$ that are not given in the form $\mathcal{O} = [Q, \mathcal{O}'\}$ for some $\mathcal{O'}$ (not $Q$-exact). As usual, the representative of a cohomology is ambiguous up to a $Q$-exact operator.

In \cite{Chang:2024zqi}, a notion to distinguish BPS cohomologies arising from a non-black-hole state versus a black-hole state has been introduced.
Consider a family of CFTs parameterized by $N$. Monotone or graviton cohomologies are the ones with a representative that is $Q$-closed and not $Q$-exact for all $N$.
On the other hand, fortuitous or black hole cohomologies are non-trivial only due to finite-$N$ effects, such as trace relations coming from the Cayley-Hamilton equations. Such relations are modified or absent as we increase $N$, so that it fails to be $Q$-closed for large $N$. 
BPS cohomologies in $\cN=4$ SYM theory are studied in many cases \cite{Chang:2022mjp, Choi:2022caq, Choi:2023znd, Choi:2023vdm, deMelloKoch:2024pcs,Gadde:2025yoa,Gaikwad:2025ugk}, also with quantum corrections \cite{Choi:2025bhi, Budzik:2025zvu, Kim:2026rnx}.

In this paper, we carry out an extensive and systematic study of BPS cohomologies of $\cN=4$ Super-Yang-Mills theory. We perform a comprehensive census for $\su{N}$ theories with $N=2, 3, 4$, mostly in the so-called BMN sector \cite{Berenstein:2002jq, Kim:2003rza}, which is closed under the classical supercharge. We find that the threshold black hole operators appear at `total charge' $j=12+6N$ or $24, 30, 36$ for $N=2, 3, 4$. The threshold black hole operators transform under the $R$-symmetry $\su{3}_R \subset \su{4}_R$ as $[3N-6, 0]$ representation, and turn out to be the brane-fused type \cite{deMelloKoch:2024pcs} (for $N\geq3$). This is holographically dual to a Dual Dressed Black Hole \cite{Choi:2024xnv, Choi:2025lck} with an $\su{2}$ `core' and $N-2$ D-branes wrapped around the core. We also compute the leading-order quantum corrections to a family of brane-fused type black hole cohomologies, and find that they are not lifted.  

It has been suspected for the past few years that all graviton operators in $\mathcal{N}=4$ SYM are of the `Coulomb type,' meaning that one can take gauge-adjoint letters to be diagonal.
This graviton Coulombness assumption was used to predict the threshold black hole operators \cite{Choi:2023vdm}. However, it was shown on multiple occasions that this assumption fails, such as in $\mathcal{N}=1$ or $\mathcal{N}=2$ theory or even $\mathcal{N}=4$ theory with $\so N$ gauge group \cite{Chang:2025mqp, Choi:2025pqr, Choi:2026faq}, or in the 3d ABJ vector model \cite{Kim:2025vup}, where examples of `non-Coulomb' type monotones, which vanish upon specializing to diagonals, have been found. 
In this paper, we also find such non-Coulomb type gravitons in $\su{3}$ and $\su{4}$ $\mathcal{N}=4$ SYM theory and explicitly write the representatives. This clearly demonstrates that generic graviton operators can be non-Coulomb type. 

One crucial organizing principle we use is the charge conjugation $\mathcal{C}$ symmetry. The $\mathcal{C}$ action turns out to commute with the supercharge, so that each charged sector in the BPS cohomology can be decomposed into $\mathcal{C}$-even and $\mathcal{C}$-odd subsectors that we call the charge conjugation parity. Interestingly, all the graviton states in the fixed charge sector have a definite charge parity. 
Therefore, we can write $H_c = H_c^{gr} \oplus H_c^{gf}$, where $H_c$ is the cohomology with charge $c$ (collectively) and $H_c^{gr}$ is its subspace with gravitons. The remaining $H_c^{gf}$ is the subspace that is free of gravitons. 
We find that all the fortuitous (or black hole) operators carry opposite parity under the charge conjugation at least up to the order we computed. 
In particular, there is one case in $\su{3}$ where a putative classical black hole cohomology at $j=36$ with no $R$-charge carries the same $\mathcal{C}$-parity as the gravitons. This could have been thought of as a new `core' black hole in $\su{3}$ theory. However, it turns out the 1-loop correction lifts this cohomology. This motivates us to conjecture a selection rule that we refer to as the $\cC$-parity conjecture.
\begin{center}
    \emph{The parity of a fortuitous cohomology is opposite to that
    of gravitons with the same charges.}
\end{center}
In other words, all fortuitous cohomologies are in the graviton-free sector $H_c^{gf}$. If it holds, it allows us to zoom into the graviton-free sector for the search of black hole cohomologies. It would be very interesting to show whether our conjecture is true, and understand the underlying physics if it holds.

The organization of this paper is as follows. In section \ref{sec:concepts}, we set up the problems and introduce some of the key ideas. In section \ref{sec:census}, we summarize the results of the BPS cohomology counting computation for $\su{2}$, $\su{3}$ and $\su{4}$. In section \ref{sec:branefused}, we study brane-fused black hole operators in detail. 

\noindent {\bf{Note Added:}}
While preparing this manuscript, we became aware of two independent groups
\cite{Sam1,Kibe:2026qid,DSD} whose results overlap with ours. We thank them for sharing their results and coordinating the submission.

\section{Concepts}\label{sec:concepts}

In this section, we set up the problems and introduce ideas that are used throughout this paper.
We first introduce the classical cohomology problem in the $\cN=4$ Yang-Mills theory and also the fortuity program.
We then introduce a systematic way developed in \cite{Choi:2023vdm} to write an ansatz for a fortuitous cohomology,
based on the idea that it should be $Q$-closed due to the trace relations.
We also introduce and further develop the idea of brane-fused operators \cite{deMelloKoch:2024pcs}, which provides a shortcut to prove non-exactness of a closed operator. 
Moreover, it allows a holographic interpretation of such operators as a black hole wrapped by branes, or Dual (giant-graviton) Dressed Black Holes.
We finally introduce the charge conjugation symmetry \cite{Beisert:2004ry,Zwiebel:2011wa,Budzik:2023vtr,Gomis:2025gzb} that serves as an extra organizing principle that also provides useful selection rules.
Although we will try to provide a self-contained introduction, the exposition will be concise.

\subsection{The fortuity program in $\cN=4$ SYM}\label{sec:setup}

Out of 16 Poincar\'e supercharges $Q^i_\alpha$, $\overline{Q}_{i\dot\alpha}$ in 4d $\cN=4$ Yang-Mills theory,
we choose, without loss of generality, to preserve $Q\equiv Q^4_-$ as well as its conjugate conformal supercharge $S = Q^\dagger$ in radial quantization.
Local gauge-invariant operators that are annihilated by the pair of supercharges are thus $\frac{1}{16}$-BPS.
By Hodge theorem, each such BPS operator represents a distinct $Q$-cohomology class, \emph{i.e.}, a set of operators that are $Q$-closed and modulo $Q$-exact operators \cite{Kinney:2005ej}.
Following the pioneering work of \cite{Grant:2008sk, Chang:2013fba}, we shall focus on the cohomology classes rather than the BPS operators themselves.
We will sometimes simply refer to their representatives as operators, with an understanding that two operators related by a $Q$-exact operator should be considered equivalent.

In the free theory limit, the $\frac{1}{16}$-BPS local operators are the gauge-invariants
constructed out of the $\frac{1}{16}$-BPS letters: 3 scalars $\phi^m$, 3 fermions (chiralinos) $\psi_m$,
2 more fermions (gauginos) $\lambda_{\dot\alpha}$, subject to an equation of motion\footnote{We will drop the dot for the $\su{2}_R$ spinor index hereafter to avoid notational clutter.}
$\partial^{\beta \dot\alpha} \lambda_{\dot\alpha} = 0$,
and a boson from the gauge field strength $f$, each of which transforms as an adjoint representation of the gauge group.
They are all contained in the free vector multiplet, \emph{i.e.}, $B_1 \overline{B}_1 [0;0]_1^{[0,1,0]}$ in the notation of \cite{Cordova:2016emh}.
The scalars are the superconformal primaries, and the others are obtained by acting supercharges $Q^m_+$ that commute with the preserved supercharge $Q$ on the primaries.
See Appendix B of \cite{Choi:2023znd} for the $\frac{1}{16}$-BPS contents of 4d $\cN=4$ superconformal multiplets.
Furthermore, two of the four derivatives $\partial_{\alpha}$ may act on each letter any number of times, acting as independent letters at the origin of the radial quantization.
All the gauge-adjoint elementary fields, or BPS letters, are assembled in a multi-trace structure to produce BPS operators, or words, at free level.

We then work in the weakly-coupled regime of the gauge theory, where the supercharge
is perturbatively expanded in the Yang-Mills coupling $g_{\rm YM}$ as
\be\label{fullQ}
\mathcal{Q} = Q + g_{\rm YM}^2 \, Q_1 + O(g_{\rm YM}^4)~.
\ee
At the leading interacting order in $g_{\rm YM}$,
also known as classical or tree-level,\footnote{To be
more complete, the full quantum supercharge is
$Q = Q_{\rm free} + g_{\rm YM} \, Q + g_{\rm YM}^2 \, Q_1 + O(g_{\rm YM}^3)$.
In this sense, the classical supercharge $Q$ is at half-loop level
since $H = \{ Q, Q^\dag \} \sim g_{\rm YM}^2$ is the one-loop Hamiltonian.
Once we restrict to operators constructed from BPS letters, $Q_{\rm free} = 0$.
Then it is conventional to rescale all fields by $g_{\rm YM}$ to write \eqref{fullQ},
see \cite{Kim:2026rnx}.}
the supercharge $Q$ acts on the BPS letters as
\begin{equation}\label{Q0action}
\begin{aligned}
[Q, \phi^m] &= 0~, \qquad &
\{Q, \psi_m\} &= \frac12 \epsilon_{mnl}[\phi^n, \phi^l]~, \qquad &
[Q, f] &= [\phi^m, \psi_m]~, \\
\{Q, \lambda_{\alpha}\} &= 0~, \qquad &
[Q, D_{\alpha}] &= [\lambda_{\alpha}, \,\cdot\,\}~. \qquad & &
\end{aligned}
\end{equation}
The cohomologies with respect to this classical supercharge $Q$ acting on free $\frac{1}{16}$-BPS local operators upon imposing the Gauss-law constraint (or gauge-invariance) give the classical $Q$-cohomologies. 

At the next perturbative level, the first quantum correction $Q_1$ to the
supercharge acts on a pair of letters in the operator as
\be
Q_1 (f_1 \cdots f_n) = \sum_{i<j} (-1)^\# Q_1 (f_i f_j) f_1 \cdots f_{i-1} f_{i+1} \cdots f_{j-1} f_{j+1} \cdots f_n~,
\ee
where the $(-1)^\#$ factor encodes the fermion anti-commutation sign that may arise from bringing $f_i$ and $f_j$ to the front.
The rules for the $Q_1$-action on pairs of elementary fields were studied in
\cite{Choi:2025bhi, Budzik:2025zvu}, see also equations (2.5) to (2.10) of \cite{Kim:2026rnx}.
For example, one particularly relevant one regarding the action of $Q_1$ on
a scalar and a fermion without derivatives reads
\be\label{Q1-psi-phi}
    Q_1 ( (\psi_n)^A (\phi^m)^B ) 
    = \frac{\kappa_1^2}{16} \delta^m_n f_{ACD}f_{BCE} \cdot
    \Big[\lambda_\alpha^D \cdot \lambda^{\alpha,E} \Big]~.
\ee
Here, $A,B,\cdots$ are gauge indices that run from $1$ to $N^2-1$
and $f_{ABC}$ is the structure constant of $\su{N}$ with the convention
\be
\tr{(T_A T_B)} = \kappa\delta_{AB}~, \quad
\tr{(T_A [T_B, T_C])}=\kappa f_{ABC}~,
\ee
where we take $\kappa = 2$ in this paper.
Here, $\kappa_1$ in \eqref{Q1-psi-phi} is a constant that depends on the gauge group
and the normalization $\kappa$ of the gauge components $A, B, \cdots$.
We leave $\kappa_1$ as a free parameter, as the overall normalization of $Q_1$
does not play any role in this paper.

Note that conventions in the current paper are related to those in \cite{Kim:2026rnx} by
\begin{equation}
\begin{aligned}
(\phi^m)_{\rm here} &= -i(\phi^m)_{\rm there}~, \quad &
(\psi_m)_{\rm here} &= -\frac{i}{2}(\psi_m)_{\rm there}~, \quad &
(f)_{\rm here} &= \frac{i}{2}(f)_{\rm there}~, \\
(\lambda_{\alpha})_{\rm here} &= -i (\partial_{\alpha} \lambda)_{\rm there}~, \quad &
(Q)_{\rm here} &= (Q_0)_{\rm there}~, \quad &
(\mathcal{Q})_{\rm here} &= (Q)_{\rm (2.13)\; there}~.
\end{aligned}
\end{equation}
With $\kappa = \kappa_1 = 2$ equal to the dual Coxeter number of $\su{2}$,
(2.3) and (2.7) of \cite{Kim:2026rnx} are reconciled from
\eqref{Q0action} and \eqref{Q1-psi-phi} here.

Some classical cohomologies with respect to $Q$ are paired up under the quantum correction
to stop being cohomologies with respect to the full quantum supercharge,
or, in other words, the tree-level BPS operators gain anomalous dimensions and are lifted to become non-BPS
\cite{Gadde:2025yoa,Chang:2025mqp,Choi:2025bhi,Kim:2026rnx}.
We will see that certain black hole cohomologies (namely brane-fused cohomologies) are not lifted under the 1-loop correction. 
However, we will also see that certain classical cohomologies (namely 'core' black hole cohomologies) are lifted via quantum correction. 



At the classical level, each BPS letter carries a certain set of quantum numbers
that simply add up in a composite operator.
The 4d $\cN=4$ superconformal algebra is given by $\mathfrak{psu}(2,2|4)$, whose bosonic subgroup is given as $\so{4, 2} \times \su{4}_R$.
Therefore, each state carries quantum numbers consisting of a scaling dimension $E$,
two angular momenta $J_1$ and $J_2$, and three Cartans of the $\su{4}_R$-symmetry $R_{1,2,3}$.
For an operator to preserve the chosen pair of supercharges $Q$ and $S$, it must saturate the unitarity bound $E \geq J_1 + J_2 + R_1 + R_2 + R_3$, thus $E$ is no longer an independent quantum number.
A subalgebra $\mathfrak{psu}(1,2|3)$ of the superconformal algebra commutes
with the chosen pair of supercharges $Q$ and $S$.
The $Q$-cohomologies are always organized into the representations of this commuting subalgebra.
In particular, each cohomology is organized into an irreducible representation of the commuting subgroup $\su{3}_R$ of the $\su{4}_R$ symmetry group. 
For example, the superscript (subscript) $m$ in $\phi^m$ ($\psi_m$) denotes the (anti-)fundamental
index under this $\su{3}_R$.

One quantum number that is widely used is the `total' charge
\begin{equation}\label{def-j}
    j \equiv 2(R_1+R_2+R_3)+3(J_1+J_2)~,
\end{equation}
which is the one combination symmetric under $J_{1,2}$ and $R_{1,2,3}$
that commutes with the chosen supercharge. This is the closest analog of the energy, which is positive definite. In the superconformal index, the fugacity conjugate to $j$ (often written as $t$) is the most natural expansion parameter.

At the classical level, \eqref{Q0action} demonstrates that the action of $Q$ on a single letter
always produces two letters. Therefore, the number of fields $Y$ of an operator is also
a well-defined quantum number, provided that we assign $Y=1$ to $Q$.
This is known as an ${\uu{1}}_Y$ \emph{bonus} symmetry that is preserved only at the
classical level \cite{Intriligator:1998ig, Chang:2013fba}, adding to the list of quantum numbers that we can utilize.
Every letter can be thought of as carrying $Y=1$, and the classical supercharge \eqref{Q0action}
also carries $Y=1$.

It is useful to define a combination called the BMN level \cite{Kim:2026rnx},
\be\label{def-dBMN}
d_{\rm BMN} \equiv R_1+R_2+R_3 + \frac{J_1 + J_2}{2} - Y~.
\ee
Every letter and thus operator carries $d_{\rm BMN} \geq 0$ and the classical supercharge
$Q$ preserves $d_{\rm BMN}$.
On the other hand, the first order quantum correction $Q_1$ increases $d_{\rm BMN}$ by 1
and more generally the supercharge $Q_n$ at the $n^{th}$ quantum order carries
$d_{\rm BMN} = n$ \cite{Kim:2026rnx},
so $d_{\rm BMN}$ is a valid quantum number only within the classical cohomology problem.
We summarize the field content as well as their quantum numbers in Table \ref{tab:fields}.

\begin{table}[t]
\begin{center}
\begin{tabular}{| c || c | c | c | c | c || c | c | c |}
\hline
Letter & \hspace{0.3cm}$R_{m}$\hspace{0.3cm} & \hspace{0.12cm}$R_{m+1}$\hspace{0.12cm} & \hspace{0.12cm}$R_{m+2}$\hspace{0.12cm} & $J \equiv J_1+J_2$ & $J_1-J_2$ &
\hspace{0.3cm}$j$\hspace{0.3cm} & \hspace{0.3cm}$Y$\hspace{0.3cm} & \hspace{0.1cm}$d_{\rm BMN}$\hspace{0.1cm} \\
\hline
$\phi^m$ & 1 & 0 & 0 & 0 & 0 & 2 & 1 & 0 \\
$\psi_m$ & $-\frac12$ & $\frac12$ & $\frac12$ & 1 & 0 & 4 & 1 & 0 \\
$f$ &0 & 0 & 0 & 2 & 0 & 6 & 1 & 0 \\
\hline
$\lambda_{\pm}$ & $\frac12$ & $\frac12$ & $\frac12$ & 0 & $\pm 1$ & 3 & 1 & $\frac12$ \\
$\partial_{\pm}$ & 0 & 0 & 0 & 1 & $\pm 1$ & 3 & 0 & $\frac12$ \\
\hline
\end{tabular}
\end{center}
\caption{Quantum numbers of the BPS elementary fields in and outside of the BMN subsector.
$m=1,2,3$ and the subscripts for $R$ are written modulo 3.}\label{tab:fields}
\end{table}

The fact that the BMN level $d_{\rm BMN}$ is always non-negative, as well as being preserved by
the classical supercharge $Q$ lets us divide the space of operators according to it.
In particular, operators with $d_{\rm BMN} = 0$ define the BMN subsector \cite{Berenstein:2002jq,Kim:2003rza}
of the $\cN=4$ SYM, which is closed under the classical supercharge.
Elementary fields in this subsector include scalars $\phi^m$, fermions (or chiralinoes) $\psi_m$
and the gauge field $f$, but no gauginos nor derivatives.
It is apparent from \eqref{Q0action} that this subsector is consistently separated
under the action of the classical supercharge.
In Table \ref{tab:fields}, the letters belonging to the BMN subsector are above the horizontal line. 
It is clear that within the BMN subsector,
\be
J_1 - J_2 = 0 \qquad \text{and} \qquad  3j - 6 (J_1 + J_2) = j + 4(R_1+R_2+R_3) = 6Y~.
\ee
This implies that in the BMN subsector, we only have one angular momentum $J = J_1+J_2$,
which can be exchanged with $Y$ for given $j$.

Since we truncate the derivatives, there are a finite number of letters
in the BMN subsector, which provides an amenable simplification of the $Q$-cohomology problem.
Despite its simplicity, the BMN subsector contains enough quantum states to account for
the black hole entropy as $N$ grows large, justifying the duality between the black hole \cite{Choi:2023vdm}.
For this reason, much of the progress in the fortuity program has been achieved within the BMN sector of $\cN=4$ SYM theory.
We will also mostly work within the BMN subsector, with the exception of the discussion of quantum cohomologies in Section \ref{sec:quantumF1} which forces us to go beyond the BMN sector.

Many of the $Q$-cohomologies of the $\cN=4$ Yang-Mills theory, as well as other holographic theories, can be constructed from the product of single-graviton operators.
In $\cN=4$ SYM, the single-graviton operators are given by the trace of symmetrized scalars, $u_n$, and their superconformal descendants.
Concretely, within the BMN sector, the gravitons in the $\su{N}$ gauge theory are given as  $u_k$, $v_k$ and $w_k$ with $k \leq N$, where
\begin{equation}\label{BMN-singlegrav}
    \begin{aligned}
        {u_2}^{ij} &\equiv \; \textrm{tr}  \left(\phi^{(i} \phi^{j)}\right)\ , \quad
        {{v_2}^{i}}_j \equiv \; \textrm{tr} \left( \phi^i \psi_{j}\right) - {\textstyle \frac{1}{3} }
        \delta^i_j\textrm{tr}\left(\phi^a \psi_{a}\right)\ , \quad
        {w_2}^i \equiv \; \textrm{tr} \left(f \phi^i + {\textstyle \frac{1}{2}} \epsilon^{ia_1a_2} \psi_{a_1} \psi_{a_2}\right)\ , \\    
        {u_3}^{ijk} &\equiv \; \textrm{tr} \left(\phi^{(i} \phi^{j} \phi^{k)}\right)\ , \\
        {{v_3}^{ij}}_k &\equiv \; \textrm{tr} \left(\phi^{(i} \phi^{j)} \psi_{k} \right) 
        - {\textstyle \frac{1}{4}} \delta^{i}_{k}  
        \textrm{tr} \left( \phi^{(j} \phi^{a)} \psi_{a} \right) 
        - {\textstyle \frac{1}{4}} \delta^{j}_{k}  \textrm{tr} \left( \phi^{(i} \phi^{a)} \psi_{a} \right)\ , \\
        {w_3}^{ij} &\equiv \; \textrm{tr}\left(f \phi^{(i} \phi^{j)} +\epsilon^{a_1a_2(i}\phi^{j)} \psi_{a_1} \psi_{a_2}\right)\ , \\    
        {u_4}^{ijkl} &\equiv \; \textrm{tr} \left(\phi^{(i} \phi^{j} \phi^{k} \phi^{l)}\right)\ , \\ 
        {{v_4}^{ijk}}_l &\equiv \; \textrm{tr} \left(\phi^{(i} \phi^{j} \phi^{k)} \psi_{l} \right) 
        - {\textstyle \frac{1}{5}} \delta^{i}_{l}  
        \textrm{tr} \left( \phi^{(j} \phi^{k} \phi^{a)} \psi_{a} \right) 
        - {\textstyle \frac{1}{5}} \delta^{j}_{l}
        \textrm{tr} \left( \phi^{(k} \phi^{i} \phi^{a)} \psi_{a} \right) 
        - {\textstyle \frac{1}{5}} \delta^{k}_{l}
        \textrm{tr} \left( \phi^{(i} \phi^{j} \phi^{a)} \psi_{a} \right)\ , \\
        {w_4}^{ijk} &\equiv \; \textrm{tr}\left(f \phi^{(i} \phi^{j} \phi^{k)}
        + \epsilon^{a_1a_2(i}\phi^{j} \phi^{k)} \psi_{a_1} \psi_{a_2}
        + {\textstyle \frac{1}{2}} \epsilon^{a_1a_2(i}\phi^{j} \psi_{a_1} \phi^{k)} \psi_{a_2}\right) \\
        &\equiv \; \textrm{tr}\left(f \phi^{(i} \phi^{j} \phi^{k)}
        + {\textstyle \frac{1}{2}} \epsilon^{a_1a_2(i}\phi^{j}
        \left( \phi^{k)} \psi_{a_1} \psi_{a_2} + \psi_{a_1} \psi_{a_2} \phi^{k)}
        + \psi_{a_1} \phi^{k)} \psi_{a_2} \right) \right)~.
    \end{aligned}
\end{equation}
The operators $u_{N+1}$ or higher, as well as their descendants, are excluded in the $\su{N}$ gauge
because they are not independent operators, but are generated by $u_2$, $\cdots$, $u_N$ through the trace relations.
Each $u_k$ is the superconformal primary of the multiplet\footnote{More precisely,
we are interested in the BPS gravitons, which are $\frac{1}{16}$-BPS
submultiplets $S_k$ \cite{Kinney:2005ej} under the commuting subalgebra $\mathfrak{psu}(1,2|3)$
of the short multiplets $B_1 \overline{B}_1 [0;0]_k^{[0,k,0]}$ under the
superconformal algebra $\mathfrak{psu}(2,2|4)$.}
$B_1 \overline{B}_1 [0;0]_k^{[0,k,0]}$, and its descendants can be derived systematically.
It has the holographic interpretation as the $k^{th}$ $S^5$ spherical harmonic modes of the
supergraviton spectrum in the AdS$_5$ supergravity, thus named gravitons \cite{Kim:1985ez,Gunaydin:1984fk,Kinney:2005ej}.

All graviton operators, spanned by products of the single-gravitons \eqref{BMN-singlegrav},
are automatically $Q$-closed because each single-graviton is.
For a finite rank gauge group, some graviton operators may be $Q$-exact or even zero,
decreasing the number of independent cohomologies.
It turns out that the number of (multi-)graviton cohomologies does not grow fast enough in $N$
to account for the black hole entropy \cite{Kinney:2005ej, Murthy:2020scj, Agarwal:2020zwm}.
However, we know that the BPS states counted by the superconformal index are large enough to account for the black hole entropy \cite{Cabo-Bizet:2018ehj, Choi:2018hmj, Benini:2018ywd}. 
This motivated the foundation of the fortuity program: to find cohomologies that do not belong to the gravitons.

A defining property of fortuitous cohomologies, as opposed to graviton, or monotone, cohomologies, is that any representative of the cohomology must stop being $Q$-closed once the same operator is mapped into the theory with a higher gauge group rank \cite{Chang:2024zqi}.
Note that not all the multi-graviton operators generated by \eqref{BMN-singlegrav} will have this property, as $Q$ acting on any of \eqref{BMN-singlegrav} vanishes as a matrix identity.
In other words, a fortuitous operator is the one that is $Q$-closed only because of a trace relation
at a finite $N$, instead of by being (cohomologous to) a multi-graviton operator that is
$Q$-closed for any $N$.
Fortuitous cohomologies can potentially represent microstates of black holes in gravity,
but they appear more sporadically and evade systematic treatment and thus have become an active research topic in recent years.

Reflecting on the definition of the fortuitous cohomologies, it is crucial to first count and list
all graviton cohomologies for finite $N$. 
There has been progress in this direction over the last several years
\cite{Chang:2022mjp,Choi:2022caq,Choi:2023znd,Choi:2023vdm,Gaikwad:2025ugk}.
Among these works, \cite{Choi:2023znd,Choi:2023vdm,Gaikwad:2025ugk} were based on the \emph{graviton Coulombness assumption} to boost the progress significantly.
The assumption is that a graviton operator is $Q$-exact, including being zero, if and only if the operator vanishes when each gauge adjoint letter is substituted by a (traceless) diagonal matrix,
\emph{i.e.}, on the Coulomb branch.

One direction of the assumption is trivial.
As every action of the supercharge \eqref{Q0action} produces a commutator,
a $Q$-exact operator cannot survive on the Coulomb branch.
The other direction, however, remains unproven and is left as an assumption.

Meanwhile, recent progress in other gauge theories besides $\su{N}$ $\cN=4$ SYM revealed that (suitably adapted version of) the graviton Coulombness assumption does not hold there, thus raising doubts.
See \cite{Gadde:2025yoa,Chang:2025mqp,Choi:2025bhi} for the 4d $\cN=4$ Yang-Mills theory with gauge groups $\so{N}$ and $\Sp{N}$, \cite{Kim:2025vup} for the ABJ vector model, \cite{Choi:2025pqr} for the 4d $\cN=4$ Yang-Mills theory with relevant deformation, and \cite{Choi:2026faq} for the Klebanov-Witten theory.
On this note, we revisit the counting of monotone and fortuitous cohomologies with gauge groups $\su{2}$, $\su{3}$ and $\su{4}$, in section \ref{sec:census}.
We indeed find that the assumption is not true in the $\cN=4$ Yang-Mills with gauge group $\su{N}$ where $N \geq 3$.

\subsection{The ansatz for non-graviton cohomologies}\label{sec:ansatz}

Recall that graviton cohomologies are generated as polynomials of a set of
single-trace operators known as single-gravitons, e.g. \eqref{BMN-singlegrav}.
We have also seen that, for gauge groups with finite rank, the graviton cohomology is not freely generated.
Due to the trace relations between finite-sized matrices,
some multi-graviton operators become $Q$-exact, disqualifying them as a cohomology.
We refer to such cohomological relations between graviton operators as graviton trace relations.
One can find a set of \emph{fundamental} graviton trace relations $\{ R_i \}$ that source
all such relations, in a sense that any relation can be written as a linear combination
(with graviton polynomials as coefficients) of them.
Within the BMN subsector where the set of single-gravitons \eqref{BMN-singlegrav} is finite,
the set of fundamental graviton trace relations is also finite.

For each fundamental graviton trace relation $R_i$, which is a $Q$-exact operator
due to trace relations,
one can find its pre-image $r_i$ such that
\be
R_i \overset{!}{=} Q (r_i)~,
\ee
where by $\overset{!}{=}$ we emphasized that the equality holds only due to trace relations.
Although it is generally a very heavy computation to find a pre-image of an operator,
it is worth the suffering to find the finite set of $r_i$'s once and for all,
because they can be used to generate all non-fundamental graviton trace relations to arbitrary order.
That is, by the definition of fundamental graviton trace relations, any graviton trace relation
(i.e. a multi-graviton operator that is $Q$-exact) for which we use the letter $s$, can be written as
\be\label{def-spoly}
s_j = \sum_i f_{i,j} (u,v,w) \, R_i \overset{!}{=} 
\sum_i f_{i,j} (u,v,w) \, Q(r_i) = 
Q \left[ \sum_i (-1)^{|f|} \, f_{i,j} (u,v,w) \, r_i \right]~,
\ee
where $f_{i,j}(u,v,w)$ are arbitrary graviton polynomials and $|f|$ its Grassmann parity.
For the last equality, we used the fact that graviton polynomials are automatically $Q$-closed.
The rightmost expression means that with $r_i$'s, we can easily construct
the pre-image of an arbitrary graviton trace relation $s_j$.
This also implies that the pre-image $r_i$, unlike $R_i$, is not a graviton,
because otherwise the entire expression \eqref{def-spoly} would vanish.

Non-graviton, or fortuitous, operators are $Q$-closed only by a trace relation
under the finite gauge group.
The relation \eqref{def-spoly} hints at a very natural candidate for such operators.
Suppose that we find a relation between (non-fundamental) graviton trace relations $s_j$,
\emph{i.e.}, $\sum_j c_{j,k} s_j = 0$.
Such a combination, a relation between relations, is called a syzygy.
Then, using \eqref{def-spoly},
\be\label{def-syzygy}
0 = \sum_j c_{j,k} s_j = \sum_{i,j} c_{j,k} \, f_{i,j} (u,v,w) \, R_i \overset{!}{=}
Q \left[ \sum_{i,j} (-1)^{|f|} \, c_{j,k} \, f_{i,j} (u,v,w) \, r_i \right]~.
\ee
Inside the bracket on the RHS is precisely a linear combination of non-graviton operators $r_i$
whose $Q$-image vanishes due to trace relations.
Thus, it is an ansatz for a fortuitous operator, pending proof of its non-$Q$-exactness.
This strategy is especially useful when one targets a specific charge sector to construct a
non-graviton cohomology, because one can construct graviton trace relations $s_j$
in that charge sector systematically and exhaustively from the fundamental $R_i$.
The number of possible $s_j$ can be easily computed using group theory,
as we will see later in this section.

Suppose that we want to construct a fortuitous operator $\cO$, within the BMN subsector,
with specific total charge $j$, angular momentum $J = J_1 + J_2$ ($J_1=J_2$ within the BMN subsector)
and representation $P$ under the global symmetry group $\su{3}_R$,
the commutant of the R-symmetry group with the supercharge $Q$.
The number of fields $Y = \frac{j}{2}-J$ is also fixed by $j$ and $J$.
We would look for graviton trace relations $s_j$ with the same quantum numbers as those of $Q \cO$.
Note that the action of supercharge $Q$ only increases the number of fields by $1$
(equivalently, decreases $J$ by $1$)
while preserving the total charge $j$ and the global symmetry representation $P$.
Given the finite list of fundamental graviton trace relations $\{ R_i \}$,
each with a definite set of quantum numbers, one can determine appropriate (tensor) products
between graviton polynomials and $\{ R_i \}$, i.e. $s_j = \sum_i f_{i,j} (u,v,w) \, R_i$,
so that the resulting $s_j$ would have the desired quantum numbers.
Once we find syzygies \eqref{def-syzygy} between $s_j$, it is straightforward to replace $R_i$ by $r_i$
inside the expressions for $s_j$, to write an ansatz for the fortuitous cohomology $\cO$.
Then, it remains to show that $\cO$ is not $Q$-exact,
and that it is not a graviton cohomology,
because even though it is $Q$-closed by trace relations, it is possible that
it is cohomologous to a graviton operator, again due to trace relations.

\subsection{Brane-fused black hole operators}\label{sec:branefusedintro}

Given a gauge-invariant operator written in the multi-trace form,
one may limit the gauge orientations of the fields to a subgroup of the gauge group.
For example in the $\su{3}$ theory, instead of treating each field as a full $\su{3}$ adjoint matrix with
8 basis matrices $T^A$, we may only keep components of its $\su{2} \times \uu{1}$ subgroup:
\be\label{3-2-1-decompose}
\phi = \sum_{A=1}^8 \phi^A T^A
~\xrightarrow{~\su{2} \times \uu{1}~}~
\sum_{A=1}^3 \phi^A T^A + \sqrt{3} \, \hat\phi \, T^8~,
\ee
or in matrix form,
\be\label{3-2-1-decompose-mat}
    \phi ~\xrightarrow{~\su{2} \times \uu{1}~}~
    \begin{pmatrix} \phi |_{\su{2}} & 0 \\ 0 & 0 \end{pmatrix} +
    \hat\phi \begin{pmatrix} 1 & 0 & 0 \\ 0 & 1 & 0 \\ 0 & 0 & -2 \end{pmatrix}~,
\ee
where we mark the $\uu{1}$ (the eighth) components with hats.
A multi-trace operator can be algebraically thought of as a polynomial in the
8 gauge components for each field, such as $\phi^A$ with $A = 1, \cdots, 8$.
The restriction \eqref{3-2-1-decompose} or \eqref{3-2-1-decompose-mat}
can be simply viewed as turning off
the off-block-diagonal components such as $\phi^{4,5,6,7}$ in the polynomial.
We shall take this restriction for the three scalars $\phi^m$ and three fermions $\psi_m$,
while for $f$ we will additionally turn off the $\uu{1}$ component $f^8 \sim \hat{f}$.

For this introduction, let us focus on the $\su{3} \to \su{2} \times \uu{1}$ restriction
for operators in the $\su{3}$ theory.
It is conceptually straightforward to generalize to other gauge groups,
e.g. $\su{4} \to \su{3} \times \uu{1}$ or even sequentially
$\su{4} \to \su{2} \times \uu{1} \times \uu{1}$, as we will in fact do in Section \ref{sec:2brane}.

Upon the restriction, an operator will become a sum of many terms,
each corresponding to a tuple of choices between which part ($\su{2}$ or $\uu{1}$),
i.e. which of the two terms in \eqref{3-2-1-decompose} or \eqref{3-2-1-decompose-mat},
to take for each field.
Since the abelian $\uu{1}$ components evade the traces, the $\su{3}$ operator will now
become a combination of $\su{2}$ operators with $\uu{1}$ components acting as
numerical coefficients.
For instances, (note that $\tr{[\phi^{k}]}_{\su{2}} = 0$ and
$ \tr{[\phi^{(i}\phi^j\phi^{k)}]}_{\su{2}} = 0$)
\begin{align}\label{u2u3decompose}
\begin{split}
 \tr{[\phi^{(i}\phi^{j)}]}
 &~\xrightarrow{~\su{2} \times \uu{1}~}~
 \tr{[\phi^{(i}\phi^{j)}]}_{\su{2}} + 6 \hat\phi^{i} \hat\phi^{j}~, \\
 \tr{[\phi^{(i}\phi^j\phi^{k)}]}
 &~\xrightarrow{~\su{2} \times \uu{1}~}~
 3\hat\phi^{(i} \tr{[\phi^j\phi^{k)}]}_{\su{2}}
 - 6\hat\phi^{(i} \hat\phi^{j} \hat\phi^{k)}~.
\end{split}
\end{align}

The restriction is entirely about the gauge group, which commutes with the global symmetry
group $\su{3}_R$. Thus, the operator after the restriction should behave as the same
representation under the global symmetry as the original operator.
Schematically, for an operator with a definite irreducible representation $R$,
\begin{equation}\label{SU2U1-intoRs}
  \cO^{(3)} [R] ~\xrightarrow{~\su{2} \times \uu{1}~}~
  \sum_{R_1 \otimes R_2 \supset R} \hat\cO [R_1] \, \cO^{(2)} [R_2]~,
\end{equation}
where each summand on the right-hand side is a product between
a $\uu{1}$ operator with some representation $R_1$ and an $\su{2}$ operator with $R_2$,
appropriately tensor producted into the representation $R$.

Importantly, the restriction preserves the $Q$-exactness property \cite{deMelloKoch:2024pcs}.
Therefore, if the original $\su{3}$ operator is $Q$-exact, i.e.
$\cO^{(3)} [R] = Q {\scriptstyle\mathcal{O}}^3 [R]$,
then the entire right-hand side of \eqref{SU2U1-intoRs} must also be a $Q$-exact operator.
Since $Q$ commutes with the global symmetry, and since it does not act on $\uu{1}$ components,
it follows that each and every $\su{2}$ operator $\cO^{(2)} [R_2]$ that appears in the sum is $Q$-exact.
Conversely, if one finds one operator $\cO^{(2)} [R_2]$ in the summation that is not $Q$-exact,
it proves that the original operator $\cO^{(3)} [R]$ cannot be $Q$-exact.

In practice, after turning off the off-block-diagonal components
from the algebraic form of the $\su{3}$ operator,
we collect the terms with certain powers of $\uu{1}$ components.
Let us take an example from \cite{deMelloKoch:2024pcs},
where the $\uu{1}$ components of the fermions $\hat\psi_m$ were also turned off, 
so that the only $\uu{1}$ components present are those of the scalars.
For an $\su{3}$ operator $(\cO^{(3)})^{ijk}$ with $R = [3,0]$, terms with cubic powers of $\hat\phi^m$
(where $m$ is the $\su{3}_R$ fundamental index) are collected.
This corresponds to restricting $R_1 = [3,0]$ since the product between bosonic variables
$\hat\phi^m$ is necessarily symmetric. As a result, one obtains an $\su{2}$ operator
$\tilde\cT^{(ijk)}_{(abc)}$ with 3 upper and 3 lower indices:
\be\label{decompose30-1}
(\cO^{(3)})^{ijk} ~\xrightarrow{~\su{2} \times \uu{1}~}~
\hat\phi^{a} \hat\phi^{b} \hat\phi^{c} \tilde\cT^{(ijk)}_{(abc)}
+ \text{(other powers of $\hat\phi$)}~.
\ee
$\tilde\cT^{(ijk)}_{(abc)}$ transforms under
$[3,0] \otimes [0,3] = [3,3] \oplus [2,2] \oplus [1,1] \oplus [0,0]$,
which are precisely the representations that may appear as $R_2$ in \eqref{SU2U1-intoRs}.
The reducible tensor $\tilde\cT$ can be written in terms of the irreducible ones.
Algebraically,
\be\label{30x03}
\tilde\cT^{(ijk)}_{(abc)} = \cT^{(ijk)}_{(abc)} + \delta^i_a \cT^{(jk)}_{(bc)}
+ \delta^{(ij)}_{(ab)} \cT^{k}_{c}  + \delta^{(ijk)}_{(abc)} \cT~,
\ee
where $\cT$'s without tilde represent irreducible representations, thus they are traceless.
For example, $\delta^a_j \cT^{(ijk)}_{(abc)} = 0$.
The four pieces of $\cT$ on the RHS correspond to the $\su{2}$ operators $\cO^{(2)} [R_2]$
in \eqref{SU2U1-intoRs}, each with appropriate $R_2$.
To prove the non-$Q$-exactness of $\cO^{(3)}$, one only needs a single summand of
\eqref{SU2U1-intoRs} that is not $Q$-exact.
Let us choose $R_2 = [0,0]$, i.e. extract the last term $\cT$ of \eqref{30x03}, which is a singlet,
from the reducible tensor $\tilde\cT^{(ijk)}_{(abc)}$.
The only covariant and equivalent way to write a singlet with the latter tensor is,
of course, the full contraction $\cT = \tilde\cT^{(ijk)}_{(ijk)}$.
If this $\su{2}$ operator is not $Q$-exact, it proves that the original $\su{3}$ operator $(\cO^{(3)})^{ijk}$
is also not $Q$-exact.

In \cite{deMelloKoch:2024pcs}, the entire fermionic tower $F_1$ of non-graviton cohomologies were
constructed using the ansatz introduced earlier and proved to be not $Q$-exact using this technique.
For each $n \geq 2$, $Q$-closed non-graviton operators with total charge $j=24+2n$ and
global symmetry representation $[n,0]$ were written down using the ansatz of section \ref{sec:ansatz}.
(For $n=1$, the ansatz does not yield any operator.)
It was explicitly constructed for $n=2,3,4$, then for $n \geq 5$ the operators follow recursively
from those with $n=3,4$ by multiplying $u$-type gravitons.
Then for these operators, the corresponding $R_2$-singlet part $\cT$ was computed.
It was zero for $n=2$, and for all $n \geq 3$, they turned out to be non-zero multiples of the smallest,
$\su{3}_R$-singlet non-graviton cohomology in the $\su{2}$ gauge theory with total charge $j=24$
found in \cite{Chang:2022mjp,Choi:2022caq}.
This result precisely matched the presence of the fermionic tower $F_1$.

In what follows, we will use the same concept in a more general way than in \cite{deMelloKoch:2024pcs}.
Namely, we will allow for fermionic $\uu{1}$ components $\hat\psi_m$, and collect terms that also include some factor of it.
Equivalently, we will choose $R_1$ in \eqref{SU2U1-intoRs} to have a non-vanishing second Dynkin label.
Although the essential principles work in the same way, there are subtleties and formalities
involving the more complicated $\uu{1}$ components.
For a streamlined explanation later, we work out in this subsection
the details of the brane-fused black hole operators where the `brane' includes a fermion.

We will be targeting operators $(\cO^{(3)})^{i_1 \cdots i_n}_j$ with global symmetry representation
$R =[n,1]$ in the $\su{3}$ gauge theory.
Consider restricting the gauge orientations as in \eqref{3-2-1-decompose}.
We still turn off the $\uu{1}$ component of the gauge field, $\hat{f}$,
but keep $\hat\phi^m$ and $\hat\psi_m$.
Then, we collect terms of $n$'th power in $\hat\phi^m$ \emph{and} linear in $\hat\psi_m$.
Algebraically, we write
\be\label{decomposen1-1}
(\cO^{(3)})^{i_1 \cdots i_n}_j ~\xrightarrow{~\su{2} \times \uu{1}~}~
\hat\phi^{a_1} \cdots \hat\phi^{a_n} \hat\psi_b \tilde\cT^{b,(i_1 \cdots i_n)}_{(a_1 \cdots a_n),j}
+ \text{(other powers of $\hat\phi$, $\hat\psi$)}~,
\ee
where $\tilde\cT$ is some set of $\su{2}$ operators, generalizing \eqref{decompose30-1}.
However, one can notice that the $\uu{1}$ part in the RHS,
namely $\hat\phi^{a_1} \cdots \hat\phi^{a_n} \hat\psi_b$,
does not transform as the irrep $[n,1]$, but rather as $[n,0] \otimes [0,1] = [n,1] \oplus [n-1,0]$,
since the trace has not been subtracted.
So ideally, we should have written the RHS as
\bea\label{decomposen1-2}
& \left( \hat\phi^{a_1} \cdots \hat\phi^{a_n} \hat\psi_b
- \frac{n}{n+2} \cdot \delta^{(a_n}_b \hat\phi^{a_1} \cdots \hat\phi^{a_{n-1})} \hat\phi^{c} \hat\psi_c \right)
\tilde\cT^{b,(i_1 \cdots i_n)}_{(a_1 \cdots a_n),j} \\
&+ \left(  \hat\phi^{a_1} \cdots \hat\phi^{a_{n-1}} \hat\phi^{c} \hat\psi_c \right)
\frac{n}{n+2} \cdot \tilde\cT^{b,(i_1 \cdots i_n)}_{(a_1 \cdots a_{n-1} b),j}
+ \text{(other powers of $\hat\phi$, $\hat\psi$)}~.
\eea
In this way, the proper $[n,1]$ component and the $[n-1,0]$ component of the $\uu{1}$ part
are separated into each line.
From another perspective, \eqref{decomposen1-2} is an algebraic way of writing the generic equation
(repeating \eqref{SU2U1-intoRs})
\begin{equation}\label{SU2U1-rep}
  \cO^{(3)} [R] ~\xrightarrow{~\su{2} \times \uu{1}~}~
  \sum_{R_1 \otimes R_2 \supset R} \hat\cO [R_1] \, \cO^{(2)} [R_2]~,
\end{equation}
restricting to terms with $R_1 = [n,1]$ and $[n-1,0]$.
Note that the reducible tensor $\tilde\cT$ obtained in \eqref{decomposen1-1}
by simply collecting terms with certain powers of $\hat\phi$ and $\hat\psi$,
directly passes on to the first line of \eqref{decomposen1-2}.
Hence it corresponds to the partial sum over $\cO^{(2)} [R_2]$ on the RHS of \eqref{SU2U1-rep}
when restricted to $R_1 = [n,1]$.
Then the analogous partial sum restricted to $R_2 = [n-1,0]$, is obtained from
$\tilde\cT$ by taking a single contraction as in the second line of \eqref{decomposen1-2}.

For the next step, we extract the singlet part of $\tilde\cT$.
That is, we choose to focus on a single term in \eqref{SU2U1-rep} with $R_2 = [0,0]$,
because that is where the threshold $\su{2}$ black hole operator may appear to complete
the proof of non-$Q$-exactness of $\cO^{(3)}$.
In the $R_1 = [n,1]$ part, i.e. the first line of \eqref{decomposen1-2},
$\tilde\cT$ transforms as $[1,n] \otimes [n,1]$, which includes the singlet $[0,0]$.
The only covariant way (up to an overall constant factor) to write a singlet with such a tensor is,
of course, the full contraction
\be\label{singletfromcT}
\cT = \tilde\cT^{j,(i_1 \cdots i_n)}_{(i_1 \cdots i_n),j}~.
\ee
If this singlet part is not $Q$-exact, for example because it is the threshold $\su{2}$ black hole
cohomology, this will prove that the original $\su{3}$ operator $\cO^{(3)}$ is not $Q$-exact.

However, with $R_1 = [n-1,0]$, $R_2 = [0,0]$ is not allowed since $R_1 \otimes R_2$ would not
contain $R = [n,1]$. This can also be shown algebraically: the tensor
$\tilde\cT^{b,(i_1 \cdots i_n)}_{(a_1 \cdots a_{n-1} b),j}$ in the second line of
\eqref{decomposen1-2} cannot be contracted to yield a singlet,
because a contraction between one of the $i$'s and $j$ would lead to zero.
Thus, we may neglect the second line of \eqref{decomposen1-2} and extract the singlet
part of the tensor $\tilde\cT$ from the first line.
Since this tensor is obtained already in \eqref{decomposen1-1},
we do not have to worry about the trace subtraction of the $\uu{1}$ part.
Simply obtaining the tensor $\tilde\cT$ from \eqref{decomposen1-1} and extracting its
singlet part would automatically lead to the particular term in \eqref{SU2U1-rep}
with $R_1 = [n,1]$ and $R_2 = [0,0]$, which is all we need.

Actually, the previous paragraph flags another caveat, from the opposite direction.
Just like the tensor $\tilde\cT$ obtained by collecting terms of power $(\hat\phi)^n (\hat\psi)$
gave contributions to the RHS of \eqref{SU2U1-rep} with not only $R_1 = [n,1]$ but also $[n-1,0]$, 
can terms with different powers of $\hat\phi$ and $\hat\psi$ contribute to the
$R_1 = [n,1]$ part of the sum \eqref{SU2U1-rep} that we are collecting?
We now argue that the answer is yes, but that such a contribution decouples from our analysis.

Thanks to the fermion statistics between three fermionic components $\hat\psi_m$,
the power of $\hat\psi$ is bounded by $3$, and we can enumerate all possibilities.
With zero factor of $\hat\psi$, the $\uu{1}$ part $\hat\phi^{a_1} \cdots \hat\phi^{a_n}$
transforms under $[n,0]$.
With one factor, $\hat\phi^{a_1} \cdots \hat\phi^{a_n} \hat\psi_b$ transforms under
$[n,0] \otimes [0,1] = [n,1] \oplus [n-1,0]$ as we have just analyzed.
With two factors, $\hat\phi^{a_1} \cdots \hat\phi^{a_{n+1}} \hat\psi_{[b_1} \hat\psi_{b_2]}$
transforms under $[n+1,0] \otimes [1,0] = [n+2,0] \oplus [n,1]$,
and with three factors $\hat\phi^{a_1} \cdots \hat\phi^{a_n} \hat\psi_{[b_1} \hat\psi_{b_2} \hat\psi_{b_3]}$
transforms under $[n,0]$.
Thus, there are two possibilities for $R_1 = [n,1]$:
\bea
(\cO^{(3)})^{i_1 \cdots i_n}_j &\xrightarrow{~\su{2} \times \uu{1}~}&
 \left( \hat\phi^{a_1} \cdots \hat\phi^{a_n} \hat\psi_b
- \frac{n}{n+2} \cdot \delta^{(a_n}_b \hat\phi^{a_1} \cdots \hat\phi^{a_{n-1})}
\hat\phi^{c} \hat\psi_c \right)
\tilde\cT^{b,(i_1 \cdots i_n)}_{(a_1 \cdots a_n),j} \nn\\
&& + \left( \hat\phi^{a_1} \cdots \hat\phi^{a_{n+1}} \hat\psi_{[b} \hat\psi_{a_{n+1}]} \right)
{\tilde{\mathcal{S}}}^{b,(i_1 \cdots i_n)}_{(a_1 \cdots a_n),j}
+ \text{(other $R_1$)}~.
\eea
%
The true $\su{2}$ operator that corresponds to the part of the sum in the RHS of \eqref{SU2U1-rep}
with $R_1 = [n,1]$ is therefore a combination of $\tilde\cT$ and $\tilde{\mathcal{S}}$.
$\tilde\cT$ has been already obtained from \eqref{decomposen1-1},
by collecting terms with $\uu{1}$ factors $(\hat\phi)^n (\hat\psi)$.
$\tilde{\mathcal{S}}$, on the other hand, should also be obtained by collecting terms
with $\uu{1}$ factors $(\hat\phi)^{n+1} (\hat\psi)^2$.

However, since the two tensors are collected with different powers of $\uu{1}$ components
that carry different sets of quantum numbers such as $j$, $J$ and $Y$,
$\tilde{\mathcal{S}}$ and $\tilde\cT$ will carry strictly different sets of quantum numbers.
In other words, the expansion \eqref{SU2U1-rep} should be further refined by
other quantum numbers than the global symmetry representations.
Then, $\tilde{\mathcal{S}}$ and $\tilde\cT$ appear as different terms of the expansion
that must be separately $Q$-exact if the original operator $\cO^{(3)}$ were to be $Q$-exact.
Conversely, it suffices to show only that $\tilde\cT$ is not $Q$-exact regardless of
$\tilde{\mathcal{S}}$, in order to show that $\cO^{(3)}$ is not $Q$-exact.
This is also why it has been consistent to turn off $\hat\psi$ in \cite{deMelloKoch:2024pcs}
when proving the non-$Q$-exactness of brane-fused operators by looking at $R_1 = [n,0]$ terms.
Terms with some factors of $\hat\psi_m$ can also contribute to the $R_1 = [n,0]$ part
of \eqref{SU2U1-rep}, but those contributions decouple because they contain different sets
of the rest of quantum numbers.\footnote{It is also for the same reason that it is consistent
to turn off $\hat{f}$, both in the present work and in \cite{deMelloKoch:2024pcs}.}
The main takeaway moving forward is that one may simply collect terms algebraically as in
\eqref{decomposen1-1} to define a (reducible) tensor $\tilde\cT$,
extract some parts of it and use it to prove the non-$Q$-exactness of the $\su{3}$ operator.

The notion of brane-fused black hole operators we have just reviewed
bears at least two significances, one technical and one conceptual.

As we mentioned at the beginning of this section, proving non-$Q$-exactness of a $Q$-closed
operator has been the computational bottleneck for constructing fortuitous cohomologies.
The idea of a brane-fused operator, while it is not applicable to all operators,
has removed this difficulty, allowing significant progress.

We also remark that the gauge restriction such as \eqref{SU2U1-intoRs},
with the $\su{2}$ black hole cohomology appearing as a part of $\tilde\cT$,
encapsulates the picture of a black hole being wrapped by a brane.
Note that the scalar trace operator $\tr [\phi^{i_1} \cdots \phi^{i_m}]$ is understood to be
dual to the dual giant gravitons, i.e., D3-branes wrapping inside the AdS$_5$
\cite{Grisaru:2000zn,Hashimoto:2000zp}, via AdS/CFT correspondence.
Note also that a D-brane reduces the five-form flux $N$ by one inside compared to the outside.
In this sense, the relevant term on the right-hand side of \eqref{SU2U1-intoRs}
depicts an AdS$_5$ black hole operator in the $\su{2}$ gauge theory wrapped by a D-brane,
thus looking like a black hole in the $\su{3}$ theory when seen from the outside.
The gravity dual picture has been of recent interest as well; see \cite{Choi:2024xnv}.

\subsection{Charge conjugation symmetry}\label{sec:parity}

The $\cN=4$ SYM theory with gauge group $\su{N}$ has a discrete charge conjugation symmetry \cite{Beisert:2004ry,Zwiebel:2011wa,Gomis:2025gzb}. 
Let $C$ be the unitary operator implementing the charge conjugation $\cC$.
Its action on local operators is given by 
\begin{equation}
    \cC (\cO) \equiv C \cO C^{-1} \;.
\end{equation}
In particular, it acts on the adjoint fields, especially on the $\frac{1}{16}$-BPS letters, as follows
\begin{equation}
    \cC : \phi^m \to -(\phi^m)^T \;, \quad
    \psi_m \to - (\psi_m)^T \;, \quad
    f \to -f^T \;, \quad
    \lambda_{\dot{\alpha}} \to - (\lambda_{\dot{\alpha}})^T \;,
\end{equation}
where $T$ denotes the $\su{N}$ matrix transpose.
The covariant derivative transforms accordingly through the gauge field, $\cC: A_{+\dot{\alpha}} \to - (A_{+\dot{\alpha}})^T$.
Charge conjugation preserves the full superconformal algebra and is a symmetry at arbitrary coupling.

The eigenvalue of charge conjugation, which we call the \textit{charge conjugation parity} or $\cC$\textit{-parity}, is $\bZ_2$-valued, i.e., $\pm 1$. 
Since charge conjugation commutes with supersymmetry, in particular $[C,Q]=0$,
we can split $Q$-cohomology classes into two sectors: parity-even and parity-odd.
Because $\cC$-parity is defined at arbitrary coupling,
this decomposition also holds for generic quantum cohomologies with respect to $\cQ$.

Charge conjugation acts on a single-trace operator with $Y$ letters including $n_f$ fermions as
\begin{equation}
    \cC: \tr\left(L_1 L_2 \cdots L_Y\right) \to (-1)^{Y+n_f(n_f-1)/2} \tr\left(L_Y \cdots L_2 L_1\right)\;,
\end{equation}
and multi-trace generalization is straightforward.
Applying this to the chiral primary, we get
\begin{equation}
    \cC: \tr(\phi^{(i_1}\phi^{i_2}\cdots\phi^{i_n)}) \to (-1)^n \tr(\phi^{(i_1}\phi^{i_2}\cdots\phi^{i_n)}) \;,
\end{equation}
where the reversed trace equals the original one due to the symmetrization of the indices.
Thus, the chiral primary operator $u_n$ is an eigenoperator of charge conjugation with parity $(-1)^n$.
Since charge conjugation commutes with the full superconformal algebra,
all graviton cohomologies in the multiplet $B_1 \overline{B}_1 [0;0]_n^{[0,n,0]}$, 
which include superconformal descendants of the chiral primary such as $v_n$ and $w_n$
\eqref{BMN-singlegrav}, are therefore charge conjugation eigenoperators with parity $(-1)^n$. 

In the classical cohomology, the number of letters is a well-defined quantum number thanks to the $U(1)_Y$ bonus symmetry.
In particular, all graviton cohomologies in $B_1 \overline{B}_1 [0;0]_n^{[0,n,0]}$
carry $n$ letters each.
Consequently, any multi-graviton cohomology with the total number of letters $Y$
has a definite parity $(-1)^Y$.
In other words, any classical cohomology with $Y$ letters and $\cC$-parity $(-1)^{Y+1}$
is necessarily of non-graviton type.
This generalizes the argument of \cite{Choi:2023znd} that, in the $\su{2}$ gauge theory,
any cohomology with an odd number of letters is of non-graviton type.
Indeed, in the $\su{2}$ gauge theory, charge conjugation is realized as a gauge symmetry, so that all gauge-invariant operators are parity-even.
Then, a graviton cohomology with an odd number of letters would need to be parity-odd,
leading to a contradiction.
Our criterion then reduces to that of \cite{Choi:2023znd}, consistent with the fact
that all gravitons in the $\su{2}$ theory have even number of letters.

Given that all graviton operators carry a definite parity $(-1)^Y$,
it is natural to question the converse.
That is, whether all non-graviton operators carry the opposite parity $(-1)^{Y+1}$.
For example, the infinite series of core fortuitous operators $\cO_n$ in the
BMN sector of the $\su{2}$ gauge theory \cite{Choi:2023znd},
including $\cO_{24}^{(2)}$ in \eqref{theO0}, turn out to have even $\cC$-parity
whereas the number of letters is odd.
Furthermore, the infinite fermionic $F_1$ tower of brane-fused black hole operators
constructed in \cite{deMelloKoch:2024pcs} all have parities opposite with respective to $(-1)^Y$.

In the next section, we will test this question in various settings,
including in gauge groups $\su{3}$ and $\su{4}$.
In fact, we will see an example in the $\su{3}$ theory where a classical fortuitous
cohomology has the same parity as gravitons, but it lifts at the first quantum level.
A similar phenomenon has been observed in the $\so{7}$ gauge theory \cite{Gadde:2025yoa}. 
A classical fortuitous cohomology was found that has both even $\cC$-parity and an even number of letters\footnote{As in the $\su{2}$ gauge theory, charge conjugation is realized as a gauge symmetry, so all gauge-invariant operators are parity-even. Thus, all gravitons consist of an even number of letters.}, and it has been shown that this is indeed quantum lifted \cite{Choi:2025bhi}.
Based on these observations, we conjecture a selection rule on BPS cohomologies at the full quantum level, that
the parity of a fortuitous cohomology is always opposite to that of gravitons with the same charges.

\section{Revisiting BPS cohomology counting}\label{sec:census}

In this section, we count graviton and fortuitous cohomologies in the BMN subsector
of the $\cN=4$ SYM theory without the graviton Coulombness assumption,
i.e., without assuming that the vanishing of a graviton polynomial with diagonal letters
implies its $Q$-exactness.
We will prove that this assumption is actually true in the $\su{2}$
gauge theory in Section \ref{sec:SU2census},
whereas we will present counterexamples and correct census data in $\su{3}$ and $\su{4}$
gauge theories in Sections \ref{sec:SU3census} and \ref{sec:SU4census} respectively.

In the counting procedure below, we frequently need to compute the ranks of very large matrices.
To this end, we adopt the random integer method and the mod-$p$ method, whose details are given in \cite{Choi:2023vdm} and \cite{Kim:2026rnx}, respectively.
When the matrices are small enough, we instead compute their ranks exactly over $\mathbb{Q}$.
If necessary, we repeat the mod-$p$ computation with two different primes as a consistency check. We especially check with at least two primes for the black hole cohomology candidates. 

\subsection{$\su{2}$}\label{sec:SU2census}

We have performed an extensive survey of the BMN sector of the $\su{2}$ theory up to $j=42$ and found that there is no non-Coulomb graviton. In fact, this can be proven. 

Let us now prove the graviton Coulombness assumption within the BMN subsector
of the $\su{2}$ gauge theory.
One direction is trivial: a $Q$-exact operator cannot be the Coulomb-type,  
i.e., it cannot be non-zero upon substituting every gauge adjoint letter
by a traceless diagonal matrix. 
We will prove the other direction, namely that if a graviton operator is non-Coulomb
(i.e., it vanishes on diagonal letters), then it is $Q$-exact, thus not a cohomology.
Thus, for the $\su{2}$ theory, a graviton operator is $Q$-exact if and only if
it vanishes on diagonal letters.

In the BMN subsector of the $\su{2}$ gauge theory, there are a finite number of single-gravitons;
6 $u_2$'s, 8 $v_2$'s and 3 $w_2$'s, under irrep $[2,0]$, $[1,1]$ and $[1,0]$ respectively
under the $\su{3}_R$ symmetry.
Of course, the 8 $v_2$'s are Grassmann-odd. See \eqref{BMN-singlegrav}.
Every graviton operator can be understood as a polynomial in these 17 `variables',
that we refer to as a graviton polynomial.

A key step for the proof is to understand that all non-Coulomb graviton
polynomials\footnote{Throughout this paper, we often use the term `non-Coulomb graviton'
for non-Coulomb type graviton cohomologies. Here, we are referring to non-Coulomb type
\emph{graviton polynomials}, which are more general multi-graviton operators
that may or may not be $Q$-exact.} are `generated' by a finite number of generators.
Consider a set of graviton polynomials that vanish (that are non-Coulomb type) once we substitute
\be
\phi^1 \to \begin{pmatrix} x_1 & 0 \\ 0 & -x_1 \end{pmatrix}~,
\ee
and likewise for every gauge-adjoint letter.
Once a graviton polynomial is non-Coulomb type, any other polynomial obtained by
multiplying another graviton polynomial to the non-Coulomb type will also be zero.
In this sense, the non-Coulomb graviton polynomials form an ideal.
One can then look for the Gr\"obner basis of the ideal, say $\{ R_i \}$.
Once the finite set of Gr\"obner basis is found, every element of the ideal can be written as
$\sum_i f_i R_i$, where $f_i$ are some graviton polynomials.
The ideal of non-Coulomb graviton polynomials can be generated by a Gr\"obner basis
$\{ R_i \}$ of size 107.\footnote{The next two paragraphs involve more technical details
on this computation and may be skipped.}

More technically, the vanishing of graviton polynomials on the Coulomb branch
is sourced by the fact that each single-graviton in \eqref{BMN-singlegrav}
is not an entirely independent variable, but can be written on the Coulomb branch as, e.g.
\be
u^{11} = \tr \left(\phi^{1} \phi^{1} \right)
= \tr \left[ \begin{pmatrix} x_1 & 0 \\ 0 & -x_1 \end{pmatrix}^2\right]~.
\ee
Thus, one should consider a super-commutative ring that consists of 7 letter components
($x_1, y_1, z_1, p_1, q_1, r_1, f_1$) and $6+8+3=17$ components of single-gravitons.
Then one can consider an ideal, which contains all non-Coulomb graviton polynomials,
that is generated by relations such as
\be
u^{11} -  \tr \left[ \begin{pmatrix} x_1 & 0 \\ 0 & -x_1 \end{pmatrix}^2\right] = 0~.
\ee
Then one can find the Gr\"obner basis of the ideal.
If the ring was defined with an elimination ordering that prioritizes the letter components
($x_1$, $\cdots$) over single-graviton components ($u^{11}$, $\cdots$),
e.g., lexicographic or the block ordering specified below,
the subset of the Gr\"obner basis that only consists
of the single-graviton components serves as a complete basis for all relations between
graviton polynomials on the Coulomb branch.
This technique has been developed and used in \cite{Choi:2023vdm},
we refer to its Section 3 for more details.

The size and details of the Gr\"obner basis may depend on further choice of variable orderings,
but any choice is equally valid.
Upon our choice that is characterized in the language of the computer algebra software 
\texttt{Singular} \cite{DGPS} as follows:
\begin{quote}\ttfamily
LIB "nctools.lib"; \\
ring R = 0, (x1,y1,z1,f1,p1,q1,r1, c(1..8), c(9..17)), \\
\hspace*{2em}(wp(2,2,2,6,4,4,4), wp(6,6,6,6,6,6,6,6, 4,4,4,4,4,4, 8,8,8)); \\
def S = superCommutative(5,15); setring S;
\end{quote}
\noindent The ideal of non-Coulomb graviton polynomials admits a Gr\"obner basis of size 107.

It is a straightforward exercise to show that each of the 107 Gr\"obner basis elements
are $Q$-exact.
Specifically, the pre-image equations $R_i = Q (r_i)$ were solved
modulo a large prime number and lifted by rational reconstruction,
then it has been checked exactly that $Q(r_i)$ indeed equal $R_i$.\footnote{The
Gr\"obner basis of size 107 is overcomplete; only 48 of them that are quadratic
in the gravitons, already generate the ideal of non-Coulomb graviton polynomials.
So it suffices to prove that these 48 are $Q$-exact, although we did prove it
for all 107 bases.}
Then, note from the definition of the basis that every element of the ideal,
that is a non-Coulomb graviton polynomial, can be written as
\be
\sum_{i=1}^{107} f_i R_i = \sum_{i=1}^{107} f_i Q(r_i)
= \sum_{i=1}^{107} Q( (-1)^{|f_i|} f_i r_i)~,  
\ee
where for the last equality we used the fact that any graviton polynomial $f_i$ is $Q$-closed,
and $|f_i|$ is its Grassmann parity.
This completes the proof that every non-Coulomb graviton polynomial is $Q$-exact,
thus there is no non-Coulomb graviton \emph{cohomology}.

This proof does not extend to $\su{N}$ theories with $N>2$.
In fact, we will shortly see counterexamples in $N=3$ and $4$.
We also note that this proof only applies within the BMN subsector.
Outside of it, there is an infinite variety of single gravitons as well as
letters involving unbounded number of derivatives,
so showing $Q$-exactness of a finite number of operators will not suffice.

In \cite{Choi:2023znd}, the graviton cohomologies in the BMN subsector of the $\su{2}$ gauge theory
have been computed exactly and completely based on the graviton Coulombness assumption,
that is now affirmed.
Subtracting from the full index, which counts all cohomologies with the $(-1)^F$ sign
as a matrix integral and is invariant under continuous quantum corrections,
the non-graviton index for the $\su{2}$ BMN subsector has been obtained:
\be\label{SU2-BMN-index}
  Z-Z_{\rm grav}=-\frac{t^{24}}{1-t^{12}}
  \cdot \prod_{m=1}^3(1 - t^2 \xi_m) \cdot 
  \prod_{m=1}^3\frac{1}{(1-t^8 \xi_m)}~,
\ee
where $t$ and $\xi_m$ subject to $\xi_1 \xi_2 \xi_3 = 1$ are fugacities
conjugate to the total charge $j$ and three $R$-charges $R_m$. 
The last factor is interpreted as simple multiplications by an arbitrary number of $w_2$-type gravitons,
the second factor accounts for the superconformal descendants within the BMN subsector,
leaving the first factor that stands for an infinite series of fermionic core $R$-singlet
fortuitous classical cohomologies.
These were all written explicitly in (3.41) of \cite{Choi:2023znd},
and here we repeat the first of them, the threshold $\su{2}$ black hole cohomology found in \cite{Chang:2022mjp,Choi:2022caq},
\be\label{theO0}
\cO_{24}^{(2)} = \ep^{a_1 a_2 a_3}
\tr (\phi^m \psi_{a_1}) \tr (\phi^n \psi_{a_2}) \tr (\psi_m \psi_n \psi_{a_3})~,
\ee
as it will be used repeatedly for constructing brane-fused operators.
Note that we will often use the notation $\cO^{(N)}_j$ to denote landmark fortuitous
cohomologies in the $\su{N}$ theory with charge $j$.
We also note that this $\cO_{24}^{(2)}$ as well as some obtained by multiplying
graviton polynomials to it were shown to be protected at the first quantum level,
whereas the next three core cohomologies $\cO_{36}^{(2)}$, $\cO_{48}^{(2)}$ and
$\cO_{60}^{(2)}$ lift due to quantum corrections \cite{Kim:2026rnx}.

\subsection{$\su{3}$}\label{sec:SU3census}

In this subsection, we show that the graviton Coulombness assumption fails in the BMN subsector of the $\su{3}$ gauge theory, unlike in the $\su{2}$ theory.
We find explicit counterexamples, i.e., \textit{non-Coulomb gravitons}, in the $\su{3}$ theory.
In addition, we revisit the graviton counting and present the revised set of
classical non-graviton cohomologies.

We perform the BPS cohomology counting for each charge sector.
Classical cohomologies can be classified by quantum numbers such as the total charge $j$,
$\su{3}_R$ representation and the number of letters as described in Section
\ref{sec:setup} as well as the $\cC$-parity we introduced in Section \ref{sec:parity},
which provide a useful classification scheme for the counting problem
since operators in different sectors cannot mix.
The (classical) supercharge $Q$ maps operators in one sector to another,
forming a chain of sectors that should be studied together in order to investigate
$Q$-closedness and exactness properties.
To this end, we study the BPS cohomologies in all charge sectors for $j\le34$
and in some selected charge sectors at $j=36$. 
In Table~\ref{tab:su3-t36} we list the selected charge sectors at $j=36$ by their
$\su{3}_R$ representations and the number of letters $Y$.
\begin{table}[h]
    \centering
    \begin{tabular}{@{}cl@{}}
    \toprule
    $\su{3}_R$ irrep. & \multicolumn{1}{c}{$Y$} \\
    \midrule
    $[0,0]$ & $7, \ldots, 18$ \\
    $[1,1]$ & $10, 11$ \\
    $[3,0]$ & $10, 11$ \\
    $[4,1]$ & $10, 11, 12$ \\
    $[5,2]$ & $10, 11, 12, 13$ \\
    $[6,0]$ & $10, 11, 12, 13$ \\
    $[7,1]$ & $11, 12, 13$ \\
    \bottomrule
    \end{tabular}
    \caption{Selected charge sectors at $j=36$ where an extensive search for BPS cohomologies has been carried out.
    Each charge sector is labeled by its $\su{3}_R$ Dynkin label and the number of letters $Y$, 
    which together specify all quantum numbers $(R_I,J_i)$.}
    \label{tab:su3-t36}
\end{table}

In each charge sector, we compute the number of independent $Q$-cohomologies
as well as the number of graviton cohomologies in each $\cC$-parity sector.
As a result, we obtain the number of fortuitous cohomologies labeled by charge sectors.
All cohomologies --- both monotones and fortuitous separately --- organize
into multiplets under the $Q$-commuting subalgebra of the superconformal algebra.
Within the BMN subsector, the descendant structure is extremely simple:
each descendant can be acted on by three components of the supercharges $Q^m_+$,
except for the chiral primaries (e.g. $u_n$ as primaries of $B_1 \overline{B}_1$ multiplets)
that are further annihilated by some of these supercharges.
It is also reflected in the $\su{2}$ non-graviton index \eqref{SU2-BMN-index}
as the middle product factor.
Given a primary fortuitous cohomology, all of its descendants can be trivially derived. 
Thus, we only list the primaries. 
In Table~\ref{tab:su3-fortuity}, we list all primary fortuitous cohomologies
in each charge sector that we have studied.
There, we have organized the cohomologies according to their total charge $j$
and the angular momentum $J = j/2-Y$.
In this way, the tower structure first observed in \cite{Choi:2023vdm} is evident.

\begin{table}[t]
    \centering
    \begin{tabular}{|c||c|c||c|}
        \hline
        \multirow{2}{*}{\diagbox{$j$}{$J$}} & \multicolumn{2}{c||}{Fermions} & Bosons \\
        \cline{2-4}
         & ~~~~5~~~~ & ~~~~7~~~~ & ~~~~6~~~~ \\ \hline
        $30$ & $[3,0]$ & &  \\ 
        $32$ & $[4,0]$ & &  \\
        $34$ & $[5,0]$ & & $[3,1]$ \\
        $36$ & $[6,0]$ & $[0,0]$ & $[4,1]$ \\ \hline
    \end{tabular}
    \caption{All $\su{3}$ superconformal primary non-graviton cohomologies up to $j=34$
    and in selected sectors of Table~\ref{tab:su3-t36} for $j=36$.
    Each cohomology is listed as its $\su{3}_R$ Dynkin labels and organized by
    two quantum numbers $j$ and $J = j/2-Y$.}
    \label{tab:su3-fortuity}
\end{table}

In addition, we have surveyed the classical BPS cohomologies beyond the BMN subsector
up to $j=23$, and found that all of them are accounted for by gravitons;
there is no fortuitous cohomology up to $j=23$.
Note that the threshold fortuitous cohomology in the full
(without restriction to the BMN subsector) $\su{2}$ theory has $j=24$ \cite{Chang:2022mjp}.
This proves that in the full $\su{3}$ theory, the threshold fortuitous cohomology
in the sense of the smallest $j$ appears at equal or higher $j$ than in the $\su{2}$ theory.

Compared with the result of \cite{Choi:2023vdm} where it had been assumed that
non-Coulomb gravitons are $Q$-exact, the singlets at $j=24$ and $30$ are now absent.
This difference is due to \textit{non-Coulomb gravitons};
these graviton operators were thought to be $Q$-exact based on the assumption,
and thus were not excluded from their Table 1, which only listed non-graviton cohomologies
after excluding graviton cohomologies.
In other words, Table 1 of \cite{Choi:2023vdm} had listed not only non-graviton cohomologies
but also non-Coulomb graviton cohomologies.
The first example of non-Coulomb gravitons appears in the $j=24$ singlet sector,
with the representative\footnote{An alternative representative is
$\epsilon_{abc} v^{ai}_j v^{bj}_k v^{ck}_i$, 
related to \eqref{eq:su3-nc1} by a $Q$-exact operator \cite{Sam1}.}
\begin{equation} \label{eq:su3-nc1}
    \epsilon_{a_1a_2a_3}u^{a_1b}{v^{a_2}}_b{v^{a_3c}}_{d}{v^{d}}_c \;.
\end{equation}
The second example, in the $j=30$ singlet sector, has the representative
\begin{equation}\label{eq:su3-nc2}
    {v^{a}}_b{v^{b}}_c{v^{c}}_d{v^{d}}_e{v^{e}}_a \;.
\end{equation}
Note that these two singlets appeared as fundamental graviton trace relations
$R_{24}^{(0,3)}$ and $R_{30}^{(0,5)}$ in \cite{Choi:2023vdm,deMelloKoch:2024pcs}
because they vanish on the Coulomb branch,
which were believed to be $Q$-exact under the graviton Coulombness assumption.
In addition, $(R_{26}^{(0,4)})^i$ is a first-level descendant of $R_{24}^{(0,3)}$;
further descendants did not appear there because they would then involve $w$-type gravitons.
We find that the singlet cohomology at $j=24$ constructed and claimed to be a
non-graviton in (4.8) of \cite{Choi:2023vdm} is in fact cohomologous to
the non-Coulomb graviton (\ref{eq:su3-nc1}).\footnote{Up to ArXiv version v2,
\cite{Choi:2023vdm} had a typo so that the operator (4.8) as written is $Q$-exact. See v3 or later.}

Apart from the two singlets at $j=24$ and $30$, all the other black hole cohomologies
reported in \cite{Choi:2023vdm} remain valid in the revised counting, within the sectors that we have studied.
Most of them are brane-fused cohomologies \cite{deMelloKoch:2024pcs}.
Explicitly, there are two towers of brane-fused cohomologies:
a fermionic tower $F_1$ in the irrep $[n,0]$ at $j=24+2n$,
and a bosonic tower $B_1$ in the irrep $[n,1]$ at $j=28+2n$, both with $n\geq3$.
Interestingly, the first brane-fused cohomology of the $F_1$ tower at $j=30$
is the threshold non-graviton cohomology of the $\su{3}$ gauge theory, in the sense of the lowest $j$.
The fermionic $F_1$ tower of brane-fused cohomologies $[n\geq3,0]$ at $j=24+2n$
has been constructed in \cite{deMelloKoch:2024pcs},
whose quantum corrections we discuss in Section \ref{sec:quantumF1}.
In Section \ref{sec:B1tower}, we construct all non-graviton cohomologies
in the bosonic tower $B_1$, i.e. $[n\geq3,1]$ at $j=28+2n$,
as brane-fused operators.
We observe that all brane-fused cohomologies with the number of letters $Y$
have charge conjugation parity $(-1)^{Y+1}$,
which is opposite to that of gravitons with the same $Y$.
This provides a simple proof that the brane-fused cohomologies are non-gravitons.

Finally, we discuss the first singlet fortuitous cohomology at $j=36$
with charges $(R_1=R_2=R_3,J)=(\frac52,7)$, the only entry in Table \ref{tab:su3-fortuity}
that does not belong to the brane-fused towers.
While we do not write the long explicit expression for its representative operator here,
we observe that this singlet fortuitous cohomology with $Y=11$ has odd $\cC$-parity,
same as that of gravitons with the same $Y=11$.
This might seem like a counterexample to the $\cC$-parity conjecture. 
It turns out, however, that this classical cohomology is lifted due to the quantum correction $Q_1$,
i.e., $Q_1 \cO^{(3)}_{36}$ is not $Q_0$-exact, meaning that this operator is no longer
$\cQ$-closed under the full quantum supercharge $\cQ$. This computation took approximately 500 core-hours (number of CPU cores times hours) on our computer system. 
This provides strong evidence for the $\cC$-parity conjecture.

\subsection{$\su{4}$}\label{sec:SU4census}

In this subsection, we extend our study to the $\su{4}$ gauge theory. 
We study the BPS cohomologies in all charge sectors up to $j=32$ and
in selected charge sectors at $j=34,36,38,40$.
In Table~\ref{tab:su4-sector} we list the selected charge sectors by their
$\su{3}_R$ representations and the number of letters $Y$.
\begin{table}[h]
    \centering
    \begin{tabular}{@{}ccl@{}}
    \toprule
    $j$ & $\su{3}_R$ irrep. & \multicolumn{1}{c}{$Y$} \\
    \midrule
    $\leq 32$ & all & all \\
    $34$ & $[5,0]$ & $9, \ldots, 12,\ 16, 17$ \\
         & $[6,1]$ & $10, \ldots, 13, 14^*$ \\
    $36$ & $[6,0]$ & $10, \ldots, 13, 14^*$ \\
    $38$ & $[7,0]$ & $11, \ldots, 13, 14^*, 15^*$ \\
         & $[8,1]$ & $15^*$ \\
    $40$ & $[8,0]$ & $12, \ldots, 14, 15^*$ \\
    \bottomrule
    \end{tabular}
    \caption{Selected charge sectors at $j=34,36,38,40$ where an extensive search for BPS cohomologies has been carried out. (All sectors $j\leq 32$ are studied.)
    Each charge sector is labeled by its $\su{3}_R$ Dynkin label and the number of letters $Y$,
    which together specify all quantum numbers $(R_I,J_i)$.
    Asterisks denote sectors where only graviton-free sectors ($\cC$-parity opposite to gravitons) are investigated.}
    \label{tab:su4-sector}
\end{table}

As in the case of $\su{3}$ gauge theory, there are examples of non-Coulomb graviton cohomologies
in the $\su{4}$ theory.
These are the cohomologies with irrep $[2,0]$ at $j=28$ and $[3,0]$ at $j=30$
that were detected in \cite{Choi:2023vdm} and assumed to be non-gravitons.
We provide a representative of the non-Coulomb graviton with irrep $[2,0]$ at $j=28$: 
\begin{align}
  \epsilon_{abc}
  \bigl(&-80 u_3^{alm} v_2^b{}_k v_2^k{}_l v_4^{c(ij)}{}_m  
  -10 u_2^{am} u_2^{kl} v_2^b{}_k v_2^c{}_l v_3^{ij}{}_m  
  -20 u_2^{al} u_2^{k(i|} v_2^b{}_k v_2^m{}_l v_3^{c|j)}{}_m 
  -12 u_2^{am} u_2^{l(i|} v_2^b{}_k v_2^{|j)}{}_l v_3^{ck}{}_m  \nn \\ 
  &+80 u_2^{a(i|} u_3^{blm} v_2^c{}_k v_2^k{}_l v_2^{|j)}{}_m 
  -20 u_2^{a(i|} u_3^{bm|j)} v_2^c{}_k v_2^k{}_l v_2^l{}_m 
  +2 u_2^{m(i|} u_3^{akl} v_2^b{}_k v_2^c{}_l v_2^{|j)}{}_m \nn \\
  &-5 \epsilon_{def}\epsilon^{(i|kl} u_2^{ad} u_2^{be} v_2^m{}_k v_2^{|j)}{}_m v_3^{cf}{}_l 
  -60 \epsilon_{def}\epsilon^{(i|kl} u_2^{ad} u_2^{b|j)} v_2^e{}_k v_2^m{}_l v_3^{cf}{}_m \nn\\
  &+50 \epsilon_{def}\epsilon^{klm} u_2^{ad} u_3^{be(i|} v_2^c{}_k v_2^f{}_l v_2^{|j)}{}_m 
  +160 \epsilon_{def}\epsilon^{(i|kl} u_3^{adm} v_2^b{}_k v_2^e{}_m v_4^{cf|j)}{}_l \nn\\
  &+160 \epsilon_{def}\epsilon^{(i|kl} u_3^{adm} v_2^b{}_k v_3^{ce}{}_m v_3^{f|j)}{}_l \bigr) \;,
\end{align}
and the one with irrep. $[3,0]$ at $j=30$:
\begin{align}
  \epsilon_{abc}\bigl(
  &600\epsilon_{def}\epsilon^{(i|lm}u_2^{ad}v_2^{|j|}{}_{l}v_4^{ben}{}_{m}v_4^{cf|k)}{}_{n}
  +340u_2^{al}u_2^{bn}v_2^{(i|}{}_{l}v_2^{|j|}{}_{m}v_4^{cm|k)}{}_{n} 
  +16u_2^{am}u_2^{b(i|}v_2^{n}{}_{l}v_3^{c|j|}{}_{m}v_3^{l|k)}{}_{n}  \nn\\
  &+64u_2^{am}u_2^{b(i|}v_2^{|j|}{}_{l}v_3^{cn}{}_{m}v_3^{l|k)}{}_{n}
  -272\epsilon_{def}\epsilon^{(i|lm}u_2^{ad}u_2^{be}v_2^{|j|}{}_{l}v_3^{cf}{}_{n}v_3^{n|k)}{}_{m} 
  -352\epsilon_{def}\epsilon^{(i|lm}u_2^{ad}u_2^{b|j|}v_2^{|k)}{}_{l}v_3^{ce}{}_{n}v_3^{fn}{}_{m} \nn \\
  &+64\epsilon_{def}\epsilon^{(i|lm}u_2^{an}u_2^{b|j|}v_2^{d}{}_{l}v_3^{ce}{}_{m}v_3^{f|k)}{}_{n}
  +48\epsilon_{def}\epsilon^{(i|lm}u_2^{ad}u_2^{n|j|}v_2^{|k)}{}_{l}v_3^{be}{}_{m}v_3^{cf}{}_{n}
  -16u_2^{a(i|}u_3^{bmn}v_2^{c}{}_{l}v_2^{|j|}{}_{m}v_3^{l|k)}{}_{n} \nn\\
  &+80u_2^{al}u_3^{bmn}v_2^{(i|}{}_{l}v_2^{|j|}{}_{m}v_3^{c|k)}{}_{n}
  +176u_2^{al}u_3^{bm(i|}v_2^{n}{}_{l}v_2^{|j|}{}_{m}v_3^{c|k)}{}_{n}
  +408u_2^{a(i|}u_3^{bln}v_2^{|j|}{}_{l}v_2^{|k)}{}_{m}v_3^{cm}{}_{n} \nn\\
  &-192u_2^{a(i|}u_3^{bl|j|}v_2^{n}{}_{l}v_2^{|k)}{}_{m}v_3^{cm}{}_{n}
  -32\epsilon_{def}\epsilon^{(i|lm}u_2^{ad}u_3^{be|j|}v_2^{c}{}_{l}v_2^{n}{}_{m}v_3^{f|k)}{}_{n}
  +16\epsilon_{def}\epsilon^{(i|lm}u_2^{a|j|}u_3^{bdn}v_2^{c}{}_{n}v_2^{e}{}_{l}v_3^{f|k)}{}_{m}\nn\\
  &+80\epsilon_{def}\epsilon^{(i|lm}u_2^{an}u_3^{bd|j|}v_2^{e}{}_{l}v_2^{|k)}{}_{n}v_3^{cf}{}_{m}
  -96\epsilon_{def}\epsilon^{(i|lm}u_2^{a|j|}u_3^{bd|k)}v_2^{e}{}_{n}v_2^{n}{}_{l}v_3^{cf}{}_{m}
  +228u_2^{a(i|}u_4^{blm|j|}v_2^{c}{}_{l}v_2^{n}{}_{m}v_2^{|k)}{}_{n} \nn\\
  &-24u_2^{a(i|}u_4^{bn|jk)}v_2^{c}{}_{l}v_2^{l}{}_{m}v_2^{m}{}_{n}
  -272u_3^{alm}u_3^{b(ij|}v_2^{c}{}_{l}v_2^{n}{}_{m}v_2^{|k)}{}_{n}
  -60u_3^{amn}u_3^{b(ij|}v_2^{c}{}_{l}v_2^{l}{}_{m}v_2^{|k)}{}_{n}\nn\\
  &+20u_2^{am}u_2^{bn}u_2^{(ij|}v_2^{c}{}_{l}v_2^{l}{}_{m}v_2^{|k)}{}_{n}
  -\epsilon_{def}\epsilon^{lmn}u_2^{ad}u_2^{b(i|}u_2^{e|j|}v_2^{c}{}_{l}v_2^{f}{}_{m}v_2^{|k)}{}_{n} \nn\\
  &+12\epsilon_{def}\epsilon^{(i|lm}u_2^{ad}u_2^{b|j|}u_2^{n|k)}v_2^{c}{}_{l}v_2^{e}{}_{m}v_2^{f}{}_{n} 
  \bigr)~.
\end{align}
See \eqref{BMN-singlegrav} for definitions of the single-gravitons.

After excluding the graviton cohomologies, including non-Coulomb,
we summarize the remaining non-graviton cohomologies in all charge sectors
that we have investigated, in Table~\ref{tab:su4-fortuity}.
All non-graviton cohomologies that are detected turn out to be fermionic.

\begin{table}[h]
    \centering
    \begin{tabular}{|c||c||c|}
        \hline
        \multirow{2}{*}{\diagbox{$j$}{$J$}} & Fermions & Bosons \\
        \cline{2-3}
         & ~~~~~~5~~~~~~ & \hphantom{~~~~~~5~~~~~~} \\ \hline
        $36$ & $[6,0]$ & --- \\ 
        $38$ & $[7,0]$ & --- \\
        $40$ & $[8,0] \times 2$ & --- \\ \hline
    \end{tabular}
    \caption{All $\su{4}$ superconformal primary non-graviton cohomologies up to $j=32$
    and in selected sectors of Table~\ref{tab:su4-sector} for $34 \leq j \leq 40$.
    Each cohomology is listed as its $\su{3}_R$ Dynkin labels and organized by
    two quantum numbers $j$ and $J = j/2-Y$.}
    \label{tab:su4-fortuity}
\end{table}

Up to the orders computed, we observe a fermionic tower with irrep $[n,0]$ ($n\geq6$) at $j=24+2n$. 
In addition to the tower, we have another independent copy of $[8,0]$ at $j=40$.
It is intriguing to think of these as brane-fused cohomology, and we will discuss this in Section~\ref{sec:branefused}.
We believe that there is no non-graviton cohomology at $j=34$
outside of the sectors that we have investigated,
which makes the brane-fused $[6,0]$ cohomology at $j=36$
the threshold fortuitous cohomology of the $\su{4}$ gauge theory.

We note that all non-graviton cohomologies in Table \ref{tab:su4-fortuity},
namely $t^{36}[6,0]$, $t^{38}[7,0]$ and $t^{40}[8,0]$ with $Y=13,14$ and $15$ respectively, 
have charge conjugation parity $(-1)^{Y+1}$ i.e. $+1,-1$ and $+1$, respectively. 
Thus, all non-graviton cohomologies found here have parity opposite to that of gravitons with the same $Y$.
This provides further evidence for our $\cC$-parity conjecture.

\section{Brane-fused black hole operators}
\label{sec:branefused}

With the census data at hand, we now proceed to write some of the lowest-lying
fortuitous cohomologies in the $\su{3}$ and $\su{4}$ gauge groups.
The threshold $\su{3}$ fortuitous cohomology in the sense of the smallest total charge $j = 30$,
which transforms as a $[3,0]$ under the $\su{3}_R$ global symmetry,
has been written as a brane-fused black hole operator in \cite{deMelloKoch:2024pcs}.
In fact, the infinite series of fortuitous operators with irrep $[n,0]$ and
charge $j = 24+2n$ ($n \geq 3$) has been similarly constructed.
In this section, we construct another infinite tower that consists of
bosonic fortuitous cohomologies with irrep $[n,1]$ and charge $j = 28 + 2n$ ($n \geq 3$)
as generalized brane-fused operators.
Together with the fermionic tower of \cite{deMelloKoch:2024pcs},
they account for all $\su{3}$ fortuitous cohomologies
detected in Section \ref{sec:SU3census} with charge $j < 36$,
where we expect the first core singlet fortuitous cohomology.
We also show that the fermionic brane-fused operators of \cite{deMelloKoch:2024pcs}
are protected at the first quantum level.
Finally, we construct the threshold fortuitous cohomology in the $\su{4}$ theory
with irrep $[6,0]$ and charge $j=36$,
as a brane-fused operator based on the $\su{3}$ threshold brane-fused operator.
We interpret it as the $\su{2}$ core black hole operator dressed, or wrapped,
by two layers of branes.

\subsection{Building the bosonic tower $B_1$}\label{sec:B1tower}

In this subsection we construct the $B_1$ tower of fortuitous cohomologies
in the $\su{3}$ theory.
Those operators have the total charge $j = 28+2n$ and transform as $[n,1]$
under the global symmetry $\su{3}_R$, for $n \geq 3$.
The first two entries ($n=3,4$) of this tower have appeared in
Table \ref{tab:su3-fortuity} which lists fortuitous cohomologies,
and further 9 entries ($n \leq 13$) appeared in Table 1 of \cite{Choi:2023vdm}
which lists fortuitous cohomologies as well as non-Coulomb monotones.

The procedure consists of three parts.
In the first part, we build ans\"atze for the non-graviton cohomologies with $n \leq 4$,
following section \ref{sec:ansatz}.
We restrict to ans\"atze that are written using graviton trace relations $s_j$ that
consist of $u$- and $v$-type gravitons, but not $w$-type.
As a result of this step, we will obtain operators that are $Q$-closed due to trace relations.
There are no such operators for $n \leq 2$, but multiple for $n = 3$ and $4$.
In the second part, we perform the $\su{3} \to \su{2} \times \uu{1}$ restriction on 
these operators following section \ref{sec:branefusedintro},
to show that at least one ansatz for $n=3$ and one for $n=4$, are not $Q$-exact.
Finally in the third part, we prove that brane-fused operators higher in the tower ($n \geq 5$)
can be obtained by multiplying gravitons appropriately to those obtained for $n=3,4$.

We start by listing (non-fundamental) graviton trace relations,
that we referred to as $s_j$ in the general discussion \eqref{def-spoly}.
Recall that there is a finite list of fundamental graviton trace relations,
i.e. combinations of graviton operators that are $Q$-exact due to trace relations.
All that consist of $u$- and $v$-type gravitons can be found in (A.4) of
\cite{Choi:2023vdm},\footnote{Among (A.4) of \cite{Choi:2023vdm},
$R_{24}^{3}[0,0]$, $R_{26}^{4}[1,0]$ and $R_{30}^{5}[0,0]$ are not $Q$-exact
as we discussed in the previous section, so these should not be used.}
and in \eqref{tr-R} we list 10 of them that can be used to construct operators in the $B_1$ tower.
They are organized according to the total charge $j$, global symmetry representation,
and angular momentum $J = J_1 + J_2$.
Suitable tensor products between the fundamental trace relations \eqref{tr-R}
and more graviton operators (we restrict to only $u$- and $v$-type gravitons)
can produce non-fundamental trace relations $s_j$ of desired quantum numbers.
Our goal is to construct operators with charge $j = 28+2n$ and representation $[n,1]$.
Furthermore, we anticipate that upon the $\su{3} \to \su{2} \times \uu{1}$ restriction
\eqref{decomposen1-1} and collecting terms with the $\uu{1}$ part $(\hat\phi)^n \hat\psi$,
the $\su{2}$ part $\tilde\cT$ will contain the $\su{2}$ threshold black hole operator
$\cO_{24}^{(2)}$ \eqref{theO0} which has angular momentum $J = 5$.
So the $\su{3}$ operator $\cO^{(3)}$ on the LHS of \eqref{decomposen1-1} needs to have $J = 6$.
Then since $s_j$'s should carry same quantum numbers as $Q \cO^{(3)}$, they will have $J = 5$.

\begin{table}
\begin{center}
\begin{tabular}{| c | r | r | r | r |}
\hline 
$R$ & \hspace{0.5cm} $\cO^{(3)}_{30}[1,1]$ & \hspace{1.3cm} $\cO^{(3)}_{32}[2,1]$ &
\hspace{1.3cm} $\cO^{(3)}_{34}[3,1]$ & \hspace{1.3cm} $\cO^{(3)}_{36}[4,1]$ \\
\hline\hline
$R_{12}^1[0,3]$ & & & & $(v_2)^4 \, R ~\to~2$ \\
\hline
$R_{12}^1[1,1]$ & & & & $(v_2)^4 \, R ~\to~2$ \\
\hline
$R_{12}^1[2,2]$ & & & & $(v_2)^4 \, R ~\to~6$ \\
\hline
$R_{14}^2[1,3]$ & & $(v_2)^3 \, R ~\to~5$ &
$(v_2)^2 \, v_3 \, R ~\to~24$ &
$u_2 \, (v_2)^3 \, R \to 15$ \\
& & & & $v_2 \, (v_3)^2 \, R \to 32$
\\
\hline
$R_{16}^2[0,1]$ & & & $(v_2)^3 \, R ~\to~2$ & $(v_2)^2 \, v_3 \, R ~\to~6$ \\
\hline
$R_{16}^2[1,2]$ & & & $(v_2)^3 \, R ~\to~5$ & $(v_2)^2 \, v_3 \, R ~\to~18$ \\
\hline
$R_{16}^2[2,3]$ & & & $(v_2)^3 \, R ~\to~5$ & $(v_2)^2 \, v_3 \, R ~\to~25$ \\
\hline
$R_{20}^3[1,0]$ & & $(v_2)^2 \, R ~\to~2$ &
$v_2 \, v_3 \, R ~\to~4$ &
$u_2 \, (v_2)^2 \, R ~\to~3$ \\
& & & & $(v_3)^2 \, R ~\to~2$ \\
\hline
$R_{22}^3[2,0]$ & & & $(v_2)^2 \, R ~\to~2$ & $v_2 \, v_3 \, R ~\to~4$ \\
\hline
$R_{30}^5[3,0]$ & & & $u_2 \, R ~\to~1$ & $u_3 \, R ~\to~1$ \\
\hline \hline
Total & 0 & 7 & 43 & 116 \\
\hline
\end{tabular}
\caption{\label{tab:s-count} Numbers of the non-fundamental graviton trace relations
$s^{[n,1]}_i$ suitable for the charge sectors of $\cO^{(3)}_{28+2n}[n,1]$ where $n \leq 4$,
for each possible combination of graviton species and fundamental graviton trace relations \eqref{tr-R}.}
\end{center}
\end{table}

We can obtain the number of independent $s_j$ operators for each $n$ based on these quantum numbers alone.
For example, consider the fundamental trace relation $R_{14}^2[1,3]$ which carries $j=14$ and $J=2$.
To obtain $s_j$ with charge $j=36$ (for $n=4$) and angular momentum $J=5$,
the only two possibilities are to multiply the trace relation
by $u_2 \, (v_2)^3$ or by $v_2 \, (v_3)^2$.
Each choice may produce operators with reducible representation
\begin{align}
[2,0] \otimes ( [1,1] \otimes [1,1] \otimes [1,1] )_A \otimes [1,3] &= \cdots + (15 \times [4,1]) + \cdots~, \nn\\
[1,1] \otimes ( [2,1] \otimes [2,1] )_A \otimes [1,3] &= \cdots + (32 \times [4,1]) + \cdots~,
\end{align}
where we are taking totally anti-symmetric products between identical fermionic tensors $v_2$ and $v_3$.
The products contain 15 and 32 copies of $[4,1]$ respectively, meaning that we can obtain that number
of $s_j$ operators with desired quantum numbers for $n=4$.
In this way, we enumerate all possible ways to obtain $s_j$ operators corresponding to $n \leq 4$,
and list the degeneracies in Table \ref{tab:s-count}.
Note that there is no possible way to write $s_j$ for $n=1$.

With these counts, we write all $s_j$ explicitly.
To do so in practice, we first lay out all indices of the tensors in a product, and consider
contracting the indices using $\delta$ and $\epsilon$ tensors so that $n$ upper and $1$ lower indices remain.
Some of such contractions turn out to be linearly dependent, or even self-vanishing.
We construct a sufficient number of contractions and extract linearly independent basis from them
until their number saturates the count in Table \ref{tab:s-count} predicted group theoretically.
We construct $7+43+116$ $s$ operators and give them the names $s^{[2,1]}_{1}, \cdots , s^{[2,1]}_{7}$,
$s^{[3,1]}_{1}, \cdots , s^{[3,1]}_{43}$ and $s^{[4,1]}_{1}, \cdots , s^{[4,1]}_{116}$.
Note that these operators are chosen for our convenience,
there are many other equally valid choices of the independent basis.
In Appendix \ref{app:gravtrrel}, we list some $s$ operators including all
that are eventually used in the construction to follow.

The next step is to find syzygies \eqref{def-syzygy}, i.e. the linear relations between $s_j$ operators.
This is a straightforward algebraic problem once $s_j$'s, consisting of graviton operators,
are completely expanded as polynomials of gauge components of the elementary fields.
We find that for $n=2$, the $7$ operators $s^{[2,1]}_{1}, \cdots , s^{[2,1]}_{7}$
are all independent. So there is no syzygy \eqref{def-syzygy},
and thus no ansatz for the fortuitous operator.
For $n=3$, there are $19$ syzygies between the $43$ operators $s^{[3,1]}_{1}, \cdots , s^{[3,1]}_{43}$.
They can be expressed in the chosen basis of $s$ operators as a $19 \times 43$ matrix of
rational numbers, which we do not display here.
Instead, we list only one particular syzygy that will be used later:
{\small
\begin{eqnarray}\label{syz31}
0 &=& ~
-914152(s^{[3,1]}_{1}) + 2468146 (s^{[3,1]}_{2}) - 2724892 (s^{[3,1]}_{6})
+ 171002 (s^{[3,1]}_{8}) - 1959484 (s^{[3,1]}_{9}) - 1139138 (s^{[3,1]}_{12}) \nn\\
&& + 4209083 (s^{[3,1]}_{13}) + 381609 (s^{[3,1]}_{15}) - 610948 (s^{[3,1]}_{16})
- 4802651 (s^{[3,1]}_{17}) - 1671855 (s^{[3,1]}_{18}) - 1442031 (s^{[3,1]}_{19}) \nn\\
&& + 1581135 (s^{[3,1]}_{20}) + 283041 (s^{[3,1]}_{22}) + 2620424 (s^{[3,1]}_{23})
- 94806 (s^{[3,1]}_{32}) + 106776 (s^{[3,1]}_{33}) + 135828 (s^{[3,1]}_{34}) \nn\\
&& - 121302 (s^{[3,1]}_{35}) + 111096 (s^{[3,1]}_{36}) + 199920 (s^{[3,1]}_{37})
- 97104 (s^{[3,1]}_{38}) + 237048 (s^{[3,1]}_{40})~.
\end{eqnarray}}
Note that this syzygy involves only three fundamental trace relations:
$s^{[3,1]}_{1}, \cdots, s^{[3,1]}_{24}$ involve $R_{14}^2 [1,3]$,
$s^{[3,1]}_{32}, \cdots, s^{[3,1]}_{36}$ involve $R_{16}^2 [2,3]$,
and $s^{[3,1]}_{37}, \cdots, s^{[3,1]}_{40}$ involve $R_{20}^3 [1,0]$.
For $n=4$, there are $63$ syzygies between the $116$ operators $s^{[4,1]}_{1}, \cdots , s^{[4,1]}_{116}$,
which we also do not write down completely.
Instead, we list the one that will be used later:
{\small
\begin{eqnarray}\label{syz41}
0 &=& ~
11447100 (s^{[4,1]}_{1}) + 22894200 (s^{[4,1]}_{2})
- 2035040 (s^{[4,1]}_{3}) - 144996600 (s^{[4,1]}_{5}) \nn\\
&& - 15262800 (s^{[4,1]}_{6}) + 244204800 (s^{[4,1]}_{7})
+ 190785000 (s^{[4,1]}_{9}) - 152628000 (s^{[4,1]}_{10}) \nn\\
&& - 58077812222 (s^{[4,1]}_{26}) - 3596084534 (s^{[4,1]}_{27}) 
+ 52910676936 (s^{[4,1]}_{28}) - 13988142320 (s^{[4,1]}_{30})\nn\\
&& + 32662675788 (s^{[4,1]}_{31}) - 6678272106 (s^{[4,1]}_{32})
- 10904013540 (s^{[4,1]}_{34}) - 175960044 (s^{[4,1]}_{37}) \nn\\
&& + 59145588 (s^{[4,1]}_{38}) - 20059973766 (s^{[4,1]}_{40})
+ 9478276638 (s^{[4,1]}_{41}) + 5261567268 (s^{[4,1]}_{42}) \nn\\
&& + 3844326288 (s^{[4,1]}_{43}) + 2773268036 (s^{[4,1]}_{44}) 
- 14324846600 (s^{[4,1]}_{45}) - 15472511712 (s^{[4,1]}_{46}) \nn\\
&& - 3342136886 (s^{[4,1]}_{47}) + 1970486682 (s^{[4,1]}_{48}) 
- 1078242930 (s^{[4,1]}_{50}) + 1878763464 (s^{[4,1]}_{52}) \nn\\
&& + 13130500296 (s^{[4,1]}_{53}) + 5514889472 (s^{[4,1]}_{54}) 
+ 2724263208 (s^{[4,1]}_{55}) + 880878828 (s^{[4,1]}_{56}) \nn\\
&& - 11705024634 (s^{[4,1]}_{57}) - 1703900835 (s^{[4,1]}_{82}) 
- 14085656550 (s^{[4,1]}_{84}) + 8791372800 (s^{[4,1]}_{85}) \nn\\
&& - 1459505250 (s^{[4,1]}_{86}) - 9029472480 (s^{[4,1]}_{87})
- 5616437850 (s^{[4,1]}_{89}) - 909553860 (s^{[4,1]}_{91}) \nn\\
&& - 1236286800 (s^{[4,1]}_{92}) - 3205188000 (s^{[4,1]}_{93})
+ 183153600 (s^{[4,1]}_{95}) + 101388600 (s^{[4,1]}_{97}) \nn\\
&& + 4742370000 (s^{[4,1]}_{98}) + 2629562400 (s^{[4,1]}_{99})
- 2452950000 (s^{[4,1]}_{100}) - 1887790320 (s^{[4,1]}_{101}) \nn\\
&& - 1556805600 (s^{[4,1]}_{105}) - 183153600 (s^{[4,1]}_{107})
+ 1013449920 (s^{[4,1]}_{110}) - 54946080 (s^{[4,1]}_{111})~.
\end{eqnarray}
}
This syzygy involves 6 fundamental trace relations, namely
$R_{12}^1 [0,3]$, $R_{12}^1 [1,1]$, $R_{12}^1 [2,2]$, $R_{14}^2 [1,3]$, $R_{16}^2 [2,3]$
and $R_{20}^3 [1,0]$.

From a syzygy, an ansatz for the non-graviton operator can be obtained simply by replacing
each $s_j$ by its pre-image, which in turn can be done by replacing
$R_i$ by its pre-image $r_i$ inside the definitions of $s$, see \eqref{def-syzygy}.
In addition, when the graviton prefactor $f_{i,j}(u,v,w)$ that multiplies $r_i$
is Grassmann odd, the sign must be flipped.
For \eqref{syz31} and \eqref{syz41}, that happens for terms that involve
the fundamental trace relation $R_{14}^2 [1,3]$ or $R_{16}^2 [2,3]$.
For the six fundamental trace relations that are used in \eqref{syz31} and \eqref{syz41},
namely $R_{12}^1 [0,3]$, $R_{12}^1 [1,1]$, $R_{12}^1 [2,2]$, $R_{14}^2 [1,3]$, $R_{16}^2 [2,3]$
and $R_{20}^3 [1,0]$, we list expressions for the corresponding pre-images in \eqref{tr-r}.
Note that these expressions are not unique: any $Q$-closed terms can be added to
these operators and still lead to the same $R_i$,
but a consistent choice should be made throughout.

Given the ans\"atze, one for $\mathcal{O}_{34}^{(3)}$ and
another for $\mathcal{O}_{36}^{(3)}$,
we now perform the $\su{3} \to \su{2} \times \uu{1}$ restriction \eqref{decomposen1-1}
on each and collect terms that contain $n$ factors of $\hat\phi$ and one of $\hat\psi$.
As a result we obtain the reducible tensor $\tilde\cT$ as in \eqref{decomposen1-1},
which transforms as a $[n,1] \otimes [1,n]$ tensor (see Section \ref{sec:branefusedintro}
for arguments regarding trace subtractions).
Then we further extract the R-symmetry singlet part $\cT$ via \eqref{singletfromcT}.
Then, $\cT$ is an $\su{2}$ operator with charges $j=24$ and $J=5$
(equivalently, it consists of $Y=7$ letters).
It must be $Q$-closed since the original $\su{3}$ operator before the restriction is.
It is known that the only $Q$-cohomology in this charge sector is the
$\su{2}$ threshold black hole cohomology $\cO^{(2)}_{24}$
presented in \eqref{theO0} and first found in \cite{Chang:2022mjp,Choi:2022caq},
and indeed we find that $\cT$ for each $n$ is its non-zero multiple:
\bea\label{cTproptoO0}
\mathcal{O}_{34}^{(3)} &~\to~ \cT = - 144353664 \cdot \cO_{24}^{(2)}~, \\
\mathcal{O}_{36}^{(3)} &~\to~ \cT = - 34616030400 \cdot \cO_{24}^{(2)}~.
\eea
This proves that the ans\"atze $\mathcal{O}_{34}^{(3)}$ and $\mathcal{O}_{36}^{(3)}$
at $j=34$ and $j=36$ respectively, are indeed not $Q$-exact.
Furthermore, fortuity of the $\su{2}$ singlet operator $\cT$ implies
fortuity of the $\su{3}$ operators $\mathcal{O}_{34}^{(3)}$ and $\mathcal{O}_{36}^{(3)}$,
since $\su{3} \to \su{2} \times \uu{1}$ restriction of an $\su{3}$ graviton
should produce $\cT$ that is also a graviton.
Therefore, $\mathcal{O}_{34}^{(3)}$ and $\mathcal{O}_{36}^{(3)}$ are
the fortuitous cohomologies that appeared in Table \ref{tab:su3-fortuity}.
On the other hand, although it is not a proof,
the fact that the analogous operators $[1,1]$ and $[2,1]$ at $j=30$ and $j=32$
are not written from this systematic approach, is naturally aligned with
their absences in Table \ref{tab:su3-fortuity}.

Finally, let us discuss the rest of the tower, namely at $n \geq 5$.
We will show that they arise as totally symmetric products between either $\cO_{34}^{(3)}$ or
$\cO_{36}^{(3)}$, and the scalar graviton operators $u_2$ and $u_3$,
following the logic of \cite{deMelloKoch:2024pcs}.
$u_2$ and $u_3$ carry charges $j=4$ and $6$, and transform under the global symmetry representations
$[2,0]$ and $[3,0]$ respectively, so suitable products can be taken to construct operators
at charge $j=28+2n$ and representation $[n,1]$ for any $n\geq 5$. For example, at $n=5$ we have
\be
(\cO_{38}^{(3)})^{(i_1 \cdots i_5)}_j \equiv
(u_2)^{(i_1 i_2} (\cO_{34}^{(3)})^{i_3 i_4 i_5)}_j
- \frac{2}{7} (u_2)^{a (i_2} (\cO_{34}^{(3)})^{i_3 i_4 i_5}_a \delta^{i_1)}_j ~,
\ee
where the second term is the trace subtraction required to ensure that
$(\cO_{38}^{(3)})^{(j i_2 \cdots i_5)}_j = 0$.
This operator is automatically $Q$-closed because both $u_2$ and $\cO_{34}^{(3)}$ are,
and it is not of graviton type as $\cO_{34}^{(3)}$ is not.
To prove that it is not $Q$-exact, we follow the same procedure outlined in Section \ref{sec:branefusedintro}.
Namely, we perform the $\su{3} \to \su{2} \times \uu{1}$ restriction
and collect terms with powers $(\hat\phi)^5 \hat\psi$ of $\uu{1}$ components.
That will lead to an $\su{2}$ operator, say $\tilde{\mathcal{U}}$, that transforms as
$[5,1] \otimes [1,5]$, corresponding to the $\cO^{(2)} [R_2]$ part after
setting $R=R_1=[5,1]$ in \eqref{SU2U1-intoRs}.
We can then extract the singlet part from $\tilde{\mathcal{U}}$ by contracting its indices as
$\mathcal{U} = \tilde{\mathcal{U}}^{b,(a_1 a_2 a_3 a_4 a_5)}_{(a_1 a_2 a_3 a_4 a_5), b}$.
It is clear from \eqref{u2u3decompose} that there are two possibilities:
either $u_2$ contributes $(\hat\phi)^2$ and $\cO_{34}^{(3)}$ does $(\hat\phi)^3 \hat\psi$,
or $u_2$ contribute no $\uu{1}$ factor and the entire $(\hat\phi)^5 \hat\psi$
comes from $\cO_{34}^{(3)}$.

Concretely, let
\begin{eqnarray}\label{more-B1-prove}
 (u_2)^{ij} &\xrightarrow{~\su{2} \times \uu{1}~}&
 {\rm tr}[\phi^{(i}\phi^{j)}]_{\su{2}} + 6 \hat\phi^{i} \hat\phi^{j}~, \\ \nn
 (\cO_{34}^{(3)})^{i_1i_2i_3}_j &\xrightarrow{~\su{2} \times \uu{1}~}& 
 \hat\phi^{a_1} \hat\phi^{a_2} \hat\phi^{a_3} \hat\psi_b \, \tilde\cT^{b,(i_1 i_2 i_3)}_{(a_1 a_2 a_3),j}
 + \hat\phi^{a_1} \cdots \hat\phi^{a_5} \hat\psi_b \, \tilde{\mathcal{S}}^{b,(i_1 i_2 i_3)}_{(a_1 \cdots a_5),j}
 + \text{(other powers of $\hat\phi$, $\hat\psi$)}~.
\end{eqnarray}
$\tilde\cT$ is the same tensor obtained earlier in this subsection,
whereas $\tilde{\mathcal{S}}$ is another $\su{2}$ operator that carries charges $j=20$ and $J=5$,
and it must be $Q$-closed since $\cO_{34}^{(3)}$ is.
However, it is known \cite{Chang:2022mjp} that there is no cohomology in such a charge sector in the $\su{2}$ theory,
so $\tilde{\mathcal{S}}$ must be $Q$-exact.
Corresponding to the two possibilities (two terms in the first line of \eqref{more-B1-prove}),
the $\su{3} \to \su{2} \times \uu{1}$ restriction of the product operator is
\bea
 (\cO_{38}^{(3)})^{i_1 \cdots i_5}_j ~\xrightarrow{~\su{2} \times \uu{1}~}~~&
 6\, \hat\phi^{(i_1} \hat\phi^{i_2|}
 \hat\phi^{a_1} \hat\phi^{a_2} \hat\phi^{a_3} \hat\psi_b \, \tilde\cT^{b,|i_3 i_4 i_5)}_{(a_1 a_2 a_3),j}
 - \frac{12}{7} \hat\phi^{a} \hat\phi^{a_1} \hat\phi^{a_2} \hat\phi^{a_3} \hat\phi^{(i_2}
 \hat\psi_b \, \tilde\cT^{b,|i_3 i_4 i_5|}_{(a_1 a_2 a_3),a}
 \delta^{i_1)}_j\\
 & + \hat\phi^{a_1} \cdots \hat\phi^{a_5} \hat\psi_b \left(
 {\rm tr}[\phi^{(i_1}\phi^{i_2|}] \, \tilde{\mathcal{S}}^{b,|i_3 i_4 i_5)}_{(a_1 \cdots a_5),j}
 - \frac27 {\rm tr}[\phi^{a}\phi^{(i_2|}] \,
 \tilde{\mathcal{S}}^{b,|i_3 i_4 i_5}_{(a_1 \cdots a_5),a} \delta^{i_1)}_j \right) \\
 & + \text{(other powers of $\hat\phi$, $\hat\psi$)}~.
\eea
Since $\tilde{\mathcal{S}}$ is $Q$-exact, its products with the $\su{2}$ graviton operator
are also $Q$-exact. Note that the $Q$-action commutes with all the global symmetry indices.
Thus, the second line will only add $Q$-exact terms to $\tilde{\mathcal{U}}$,
and it will not be useful for proving that $\cO_{38}^{(3)}$ is not $Q$-exact.
On the other hand, the first line is
\be
\hat\phi^{a_1} \cdots \hat\phi^{a_5} \hat\psi_b \left(
6\, \delta^{(i_1}_{(a_4} \delta^{i_2|}_{a_5} \tilde\cT^{b,|i_3 i_4 i_5)}_{a_1 a_2 a_3),j}
-\frac{12}{7} \delta^{(i_1}_j  \delta^{i_2|}_{(a_5} \tilde\cT^{b,|i_3 i_4 i_5)}_{a_1 a_2 a_3,a_4)}
\right)~.
\ee
The $[5,1] \otimes [1,5]$ tensor in the parenthesis is $\tilde{\mathcal{U}}$, whose singlet part
$\mathcal{U} = \tilde{\mathcal{U}}^{b,(a_1 a_2 a_3 a_4 a_5)}_{(a_1 a_2 a_3 a_4 a_5), b}$
we would like to extract. Taking care with symmetry factors, it gives
\begin{eqnarray}
\mathcal{U} &=& 6\, \delta^{(a_1}_{a_4} \delta^{a_2|}_{a_5} \tilde\cT^{b,|a_3 a_4 a_5)}_{a_1 a_2 a_3,b}
-\frac{12}{7} \delta^{(a_2}_{a_5} \tilde\cT^{a_1,a_3 a_4 a_5)}_{a_1 a_2 a_3,a_4} \nn\\
&=& 6 \left(
\frac{1}{20} \delta^{a_4}_{a_4} \delta^{a_5}_{a_5} \tilde\cT^{b,(a_1 a_2 a_3)}_{a_1 a_2 a_3,b}
+\frac{7}{20} \delta^{a_4}_{a_4} \tilde\cT^{b,(a_1 a_2 a_3)}_{a_1 a_2 a_3,b}
+\frac{12}{20} \tilde\cT^{b,(a_1 a_2 a_3)}_{a_1 a_2 a_3,b} \right) \nn\\
&& -\frac{12}{7} \left(
\frac15 \delta^{a_5}_{a_5} \tilde\cT^{(a_1,a_2 a_3 a_4)}_{a_1 a_2 a_3,a_4}
+ \frac45 \tilde\cT^{(a_1,a_2 a_3 a_4)}_{a_1 a_2 a_3,a_4}\right) \nn\\
&=& \frac{63}{5} \tilde\cT^{b,(a_1 a_2 a_3)}_{a_1 a_2 a_3,b}
- \frac{3}{5} \tilde\cT^{a_4,(a_1 a_2 a_3)}_{a_1 a_2 a_3,a_4} \nn\\
&=& 12 \, \cT~.
\end{eqnarray}
We used the fact that $\tilde\cT^{(a_1,a_2 a_3 a_4)}_{a_1 a_2 a_3,a_4}
= \frac14 \tilde\cT^{a_4,(a_1 a_2 a_3)}_{a_1 a_2 a_3,a_4}$~,
since $\tilde\cT$ being a $[3,1] \otimes [1,3]$ tensor means the lower index $a_4$
cannot contract with one of the three symmetrized upper indices.
$\cT = \tilde\cT^{b,(a_1 a_2 a_3)}_{a_1 a_2 a_3,b}$ is precisely the singlet part
of the tensor $\tilde\cT$, and it is
$\cT = - 144353664 \cdot \cO_{24}^{(2)}$ as we have shown above.
We observe that this singlet part, which is a non-trivial fortuitous operator in the
$\su{2}$ theory, is directly carried into the $(\hat\phi)^5 \hat\psi$ term of the
$\su{2} \times \uu{1}$ restriction of $\cO_{38}^{(3)}$,
obstructing the latter from being $Q$-exact.

It is straightforward to extend this argument to arbitrarily high $n$.
For any given $n\geq 5$, one can construct an ansatz $\cO_{28+2n}^{(3)}$ by taking the symmetric
tensor product between $u_2$ and/or $u_3$ gravitons and one of the lower $\cO_{28+2n'}^{(3)}$.
The only term with the $\uu{1}$ part $(\hat\phi)^n \hat\psi$ that is not trivially $Q$-exact
will have to arise from the $\cT$ of $\cO_{34}^{(3)}$ or $\cO_{36}^{(3)}$ in \eqref{cTproptoO0},
with $u_2$ and $u_3$ gravitons only contributing the correct number of $\hat\phi$ factors.
This term will keep being a non-zero multiple of the threshold $\su{2}$ black hole operator $\cO_{24}^{(2)}$,
proving the non-$Q$-exactness of $\cO_{28+2n}^{(3)}$.

A similar remark to that made in \cite{deMelloKoch:2024pcs} follows regarding their uniqueness.
Apparently, for sufficiently large $n$ there are multiple ways to construct $\cO_{28+2n}^{(3)}$.
For example, already for $n=6$, one can either multiply $u_3$ to $\cO_{34}^{(3)}$
or $u_2$ to $\cO_{36}^{(3)}$. The proof of non-$Q$-exactness will hold for any such possibilities.
However, this does not prove that there is more than one cohomologies for each $n$.
Because the proof boils down to showing that the corresponding $\cT$ part is a non-zero multiple of
a single $Q$-cohomology $\cO_{24}^{(2)}$,
if there are $k$ possible ways of constructing $\cO_{28+2n}^{(3)}$ as mentioned above,
there are always $k-1$ independent combinations of them for which the attempted proof leads to a vanishing $\cT$.
Therefore, our `trick' for proving the non-$Q$-exactness can only verify $k$ independent cohomologies
among a given charge sector, when there are $k$ independent `core' cohomologies in the smaller gauge group
that obstruct the $Q$-exactness of the original operators. In this case, $k$ was 1.

\subsection{$F_1$ tower is one-loop protected}\label{sec:quantumF1}

Throughout this work, we have mostly been studying the classical $Q$-cohomology
with respect to the tree-level, or classical, action $Q$ \eqref{Q0action}
of the full quantum supercharge $\mathcal{Q}$.
However, it has been recently found \cite{Gadde:2025yoa,Chang:2025mqp,Choi:2025bhi,Kim:2026rnx}
that some classical cohomologies may lift once quantum corrections to the supercharge are taken into account,
meaning that the corresponding classical BPS operators lift to non-BPS.
In this subsection, we show that the $F_1$ tower of brane-fused black hole operators
constructed in \cite{deMelloKoch:2024pcs}, are not lifted at the first quantum level,
i.e. under $Q_1$ in \eqref{fullQ}.

There are two mechanisms under which a classical, $Q$-cohomology may lift at the
first quantum level \cite{Kim:2026rnx}.
First, if the $Q_1$-image of the operator is not $Q$-exact, it is no longer $Q$-closed
at the quantum level and thus lifts.
Second, if a classical cohomology is a $Q_1$-image of another classical cohomology,
it is no longer non-$Q$-exact at the quantum level and thus lifts.
Note that the two mechanisms correspond to a pair of classical cohomologies lifting together.
As $Q_1$ increases the BMN level $d_{\rm BMN}$ by 1, the $Q_1$-image side of the pair
always has $d_{\rm BMN}$ higher by 1 than the pre-image side.
But since $d_{\rm BMN}$ is always non-negative, an operator in the BMN subsector
with $d_{\rm BMN} = 0$ cannot be a $Q_1$-image of another classical cohomology,
so it cannot lift under the second mechanism.
Therefore, it suffices to show that the $Q_1$-image of the classical cohomology is $Q$-exact,
in order to show that it is protected at the first quantum level.

We have shown, by brute force, that the $Q_1$-images of the first two brane-fused cohomologies in the $F_1$ tower,
namely the $\cO_{30}^{(3)}$ and $\cO_{32}^{(3)}$ written in \cite{deMelloKoch:2024pcs}
with representations $[3,0]$ and $[4,0]$ respectively,
are $Q$-exact. Thus, the two are protected at the first quantum level.

Then recall that the rest of the infinite tower is obtained by taking symmetric products between
either of the first two operators and any appropriate numbers of $u_2$ and $u_3$ gravitons.
We now prove that the $Q_1$-image of any such product is also $Q$-exact.

Let $\cO$ be a classical cohomology in the BMN subsector that consists of $\phi$ and $\psi$ but not $f$.
We assume that $Q_1 \cO$ is $Q$-exact, having $\cO_{30}^{(3)}$ or $\cO_{32}^{(3)}$ in mind.
Consider the product $u_{2,3} \cO$, where $u_{2,3}$ could either be $u_2$ or $u_3$.
$Q_1 \left( u_{2,3} \cO \right)$ consists of two parts.
The first is obtained by $Q_1$ acting on a pair of letters inside $\cO$, giving 
$u_{2,3} \cdot Q_1\cO$ which is $Q$-exact by assumption.
The second is obtained by $Q_1$ acting on one $\phi$ in $u_{2,3}$ and one $\psi$ in $\cO$.
There is no third part because $Q_1$ does not act on a pair of scalars.
Because of symmetry between $\phi$'s inside $u_{2,3}$, it suffices to show the $Q$-exactness
of the part of $Q_1 \left( \tr \left(\phi^{m} \Phi \right) \cO \right)$
where $Q_1$ acts on $\phi^{m}$ and a $\psi$ inside $\cO$.
$\Phi = \phi^n$ if we have $u_2$ in mind, or $\Phi = \phi^{(n} \phi^{p)}$ if we have $u_3$ in mind.
Furthermore, if $\cO$ contains multiple $\psi$'s, it will simply be a (signed) sum over the $\psi$'s,
so we can target a specific $\psi$ one by one.
Suppose that the $\psi$ that is targeted appears in $\cO$ as $\tr \left( \psi_n \Psi \right)$,
where $\Psi$ could be any composite field.
For concreteness, let $\cO = \tr \left( \psi_n \Psi \right) \cdot \tilde\cO$,
where $\tilde\cO$ simply denotes the rest of the operator.
The part that we want to show $Q$-exactness of is now written, using \eqref{Q1-psi-phi}, as
\begin{eqnarray}
Q_{1,\phi^m,\psi_n} \left( \tr \left(\phi^{m} \Phi \right) \tr \left( \psi_n \Psi \right) \tilde\cO \right)
&=&~ \frac{\kappa_1^2}{16} \delta^m_n f_{ACD} f_{BCE} \lambda_\alpha^D \lambda^{\alpha,E}
\tr \left[ \Phi T_B \right] \tr \left[ T_A \Psi \right] \tilde\cO \nn \\
&=&~ \frac{\kappa_1^2}{16} \delta^m_n \lambda_\alpha^D \lambda^{\alpha,E}
\tr \left[ T_C \, [T_E, \Phi] \right] \tr \left[ T_C \, [T_D, \Psi] \right] \tilde\cO \nn \\
&=&~ \frac{\kappa_1^2 \kappa}{16} \delta^m_n \lambda_\alpha^D \lambda^{\alpha,E}
\tr \left[ [T_E, \Phi] [T_D, \Psi] \right] \tilde\cO \nn \\
&=&~ \frac{\kappa_1^2 \kappa}{16} \delta^m_n
\tr \left[ \left\{ \lambda^\alpha , [\lambda_\alpha , \Phi] \right\} \Psi \right] \tilde\cO \nn \\
&=&~ -\frac{\kappa_1^2 \kappa}{16} \delta^m_n
\tr \left[ \left( Q [\lambda^\alpha , D_\alpha \Phi] \right) \Psi \right] \tilde\cO~.
\end{eqnarray}
In a sense, we can understand $Q_{1,\phi^m,\psi_n} \left( \tr \left(\phi^{m} \Phi \right) \cdot \cO \right)$
as an operation of replacing one $\psi_n$ in $\cO$ by
$-\frac{\kappa_1^2 \kappa}{16} \delta^m_n \cdot Q [\lambda^\alpha , D_\alpha \Phi]$.
Then we need to sum over all $\psi$'s in $\cO$, which makes it a (Grassmann-odd) functional derivative
\be\label{Q1crossterms}
Q_{1,\phi^m} \left( \tr \left(\phi^{m} \Phi \right) \cdot \cO \right)
=\left( -\frac{\kappa_1^2 \kappa}{16} \delta^m_n \cdot Q [\lambda^\alpha , D_\alpha \Phi] \right)
\frac{\delta}{\delta \psi_n} \cO~.
\ee

Now, define another operation $D_\Phi$ that replaces each $\psi$ by
$-\frac{\kappa_1^2 \kappa}{16} \delta^m_n \cdot [\lambda^\alpha , D_\alpha \Phi]$
(note the absence of $Q$ compared to \eqref{Q1crossterms}).
This is a bosonic derivative, since it is replacing a fermionic $\psi$ with another fermionic piece.
Then consider $[Q, D_\Phi] \cO$.
In the coherent language, $Q$ is an operation that replaces each $\psi_n$ by
$\frac12 \ep_{nmp} [\phi^m, \phi^p]$, but it can also act on a letter with derivatives.
Since $\cO$ consists only of $\phi$ and $\psi$, of which $Q$ only acts on $\psi$,
$Q D_\Phi \cO$ and $D_\Phi Q \cO$ are almost identical.
They would both choose an ordered pair of $\psi$'s from $\cO$, replace one $\psi_n$ with
$-\frac{\kappa_1^2 \kappa}{16} \delta^m_n \cdot [\lambda^\alpha , D_\alpha \Phi]$
and the other $\psi_r$ with $\frac12 \ep_{rmp} [\phi^m, \phi^p]$, then sum over the pairs.
There is a class of terms, however, that is included in $Q D_\Phi \cO$ but not in $D_\Phi Q \cO$.
It arises when $Q$ acts on $D_\alpha \Phi$ that is already produced by $D_\Phi$,
but not on any of the $\psi$'s in $\cO$.
Such terms can be written as
\be
[Q, D_\Phi] \cO
= Q \left( -\frac{\kappa_1^2 \kappa}{16} \delta^m_n \cdot [\lambda^\alpha , D_\alpha \Phi] \right)
\frac{\delta}{\delta \psi_n} \cO
= -\frac{\kappa_1^2 \kappa}{16} \delta^m_n \cdot Q[\lambda^\alpha , D_\alpha \Phi]
\frac{\delta}{\delta \psi_n} \cO~.
\ee
That is, $Q$ acting on the part that has been created by $D_\Phi$,
precisely restores the $Q$ factor which was the difference between \eqref{Q1crossterms} and $D_\Phi$.
Meanwhile, by assumption $\cO$ is a classical cohomology, thus $Q \cO = 0$.
Therefore, we arrive at
\be\label{Q1isQ0exact}
Q_{1,\phi^m} \left( \tr \left(\phi^{m} \Phi \right) \cdot \cO \right)
= [Q, D_\Phi] \cO = Q ( D_\Phi \cO)~,
\ee
proving that the LHS is $Q$-exact.
The true action of $Q_1$ on the product operator such as $(u_2)^a (u_3)^b \cO$
will be a sum over \eqref{Q1isQ0exact}, corresponding to each $\phi$ inside the
$(u_2)^a (u_3)^b$ part and with appropriate $\Phi$ for each,
plus the part where $Q_1$ acts entirely on $\cO$.
Summing over all such $Q$-exact contributions, we conclude that any product 
$(u_2)^a (u_3)^b \cO$ where $\cO$ is a classical cohomology that consists only of $\phi$ and $\psi$
and that is protected at the $Q_1$ level, is also protected at the $Q_1$ level.
This proves that the entire $F_1$ tower of brane-fused black hole operators is protected.

Protection of the $\su{3}$ $F_1$ tower under the first order quantum corrections
extends the observation made in \cite{Kim:2026rnx} regarding $\su{2}$ cohomologies
that cohomologies constructed by dressing a core classical cohomology
are likely to inherit the protection of the core.
It will be extremely interesting to find further (counter-)examples or
explanations of this phenomenon.

\subsection{$\su{4}$: double brane-fused operators}\label{sec:2brane}

Finally, we analyze the infinite tower of $\su{4}$ fortuitous cohomologies
with $[n \geq 6, 0]$ at charge $j = 24+2n$.
We construct them as brane-fused type operators in a sense that they contain the
$\su{3}$ $F_1$ tower of operators
under the $\su{4} \to \su{3} \times \uu{1}$ restriction.
Since the $F_1$ tower consists of brane-fused operators
that contain the $\su{2}$ threshold R-singlet fortuitous operator at $j=24$,
we interpret the tower of $\su{4}$ operators as double brane-fused operators.

The threshold fortuitous cohomology in the $\su{4}$ theory, $\cO^{(4)}_{36}$,
was detected during the census presented in Section \ref{sec:SU4census}.
It has the total charge $j=36$, angular momentum $J=5$ and number of fields $Y=13$,
and transforms under the global symmetry $\su{3}_R$ as $[6,0]$.
To extract useful information from a huge expression for this operator,
we perform an $\su{4} \to \su{3} \times \uu{1}$ gauge restriction and follow a
completely analogous procedure to \cite{deMelloKoch:2024pcs} or to Section \ref{sec:branefusedintro}.
We restrict the gauge orientations of the scalars $\phi^m$ as the $\su{4}$
generalization of \eqref{3-2-1-decompose-mat},
with tildes playing the role of hats for the new $\uu{1}$, i.e.
\begin{eqnarray}\label{4-3-1-decompose-mat}
\phi = \sum_{A=1}^{15} \phi^A T^A
    &~\xrightarrow{~\su{3} \times \uu{1}~}~&
    \sum_{A=1}^{8} \phi^A T^A + \sqrt{6}\, \tilde\phi \, T^{15}~, \nn\\
\phi &~\xrightarrow{~\su{3} \times \uu{1}~}~&
    \begin{pmatrix} \phi |_{\su{3}} & 0 \\ 0 & 0 \end{pmatrix} +
    \tilde\phi \begin{pmatrix}
    1 & 0 & 0 & 0 \\ 0 & 1 & 0 & 0 \\ 0 & 0 & 1 & 0 \\ 0 & 0 & 0 & -3
    \end{pmatrix}~.
\end{eqnarray}
For fermions $\psi_m$ and the gauge field strength $f$, we further turn off their
$\uu{1}$ components $\tilde\psi$, as well as $\tilde{f}$.
Then we collect terms in cubic power of the scalar $\uu{1}$ components
$\tilde\phi^{a_1} \tilde\phi^{a_2} \tilde\phi^{a_3}$ and read off its $\su{3}$ part
$\tilde\cT^{i_1 \cdots i_6}_{a_1 a_2 a_3}$ analogous to \eqref{decompose30-1}.
This tensor $\tilde\cT$ transforms under the global symmetry as
$[6,0] \otimes [0,3] = [6,3] \oplus [5,2] \oplus [4,1] \oplus [3,0]$, from which
we extract the $[3,0]$ part $\cT^{i_1i_2i_3} = \tilde\cT^{(i_1i_2i_3a_1a_2a_3)}_{(a_1a_2a_3)}$.

The $[3,0]$ part, namely $\cT^{i_1 i_2 i_3}$, belongs precisely to the cohomology
of the $\su{3}$ threshold operator $\cO^{(3)}_{30}$ at $j=30$,
written in \cite{deMelloKoch:2024pcs}.\footnote{$\cT$ can only be cohomologous to
$\cO^{(3)}_{30}$ or zero, since it is the only cohomology in the correct charge sector
with the parity opposite to gravitons.}
Containing the $\su{3}$ fortuitous operator in one of its expansion terms
under $\su{4} \to \su{3} \times \uu{1}$, an obvious analogue of \eqref{SU2U1-intoRs},
this proves that the $\su{4}$ threshold operator $[6,0]$ at $j=36$
is a brane-fused black hole operator.

Meanwhile, we recall that the $\su{3}$ threshold operator $\cO^{(3)}_{30}$
is itself a brane-fused operator that contains the $\su{2}$ threshold
fortuitous operator $\cO^{(2)}_{24}$ at $j=24$ \cite{Chang:2022mjp,Choi:2022caq}
upon the $\su{3} \to \su{2} \times \uu{1}$ gauge restriction.
Connecting the two results, it implies that once the $\su{4}$ threshold operator
$\cO^{(4)}_{36}$ is restricted as $\su{4} \to \su{2} \times \uu{1}_x \times \uu{1}_y$,
the $[6,0]$ operator contains, among the expansion such as in \eqref{decompose30-1},
\be\label{decompose60}
(\cO^{(4)}_{36})^{i_1 \cdots i_6} ~\xrightarrow{~\su{2} \times \uu{1}_x \times \uu{1}_y~}~
\hat\phi^{(i_1} \hat\phi^{i_2} \hat\phi^{i_3}
\tilde\phi^{i_4} \tilde\phi^{i_5} \tilde\phi^{i_6)} \cO^{(2)}_{24}
+ \cdots ~,
\ee
where $\hat\phi$ and $\tilde\phi$ refers to the $\uu{1}_x$ and $\uu{1}_y$ components
of the scalars, respectively.

We now turn to the next $\su{4}$ fortuitous cohomology, a $[7,0]$ at $t=38$,
see Table \ref{tab:su4-sector}.
We perform a similar $\su{4} \to \su{3} \times \uu{1}$ gauge restriction \eqref{4-3-1-decompose-mat},
this time collecting terms in quartic power of the $\uu{1}$ scalar.
This leads to an $\su{3}$ operator $\tilde\cT$ that transforms as 
$[7,0] \otimes [0,4]$, from which we can extract a $[3,0]$ part.
The $[3,0]$ part $\cT$ thus obtained, again belongs precisely to the cohomology of $\cO^{(3)}_{30}$.
Thus, the $[7,0]$ is also constructed as a brane-fused black hole operator
enclosing the $\su{3}$ fortuitous operator $\cO^{(3)}_{30}$
and even the $\su{2}$ operator $\cO^{(2)}_{24}$ in the sense of \eqref{decompose60}.
We choose not to include the explicit expression for this operator.

Climbing up the tower, the $\su{4}$ operators with $[n \geq 8,0]$ at $j=24+2n$,
can be constructed as symmetric products between $u$-type gravitons and $\cO^{(4)}_{36}$.
This follows from the same argument used for construction of the infinite towers
$F_1$ and $B_1$ in the $\su{3}$ theory, as we now explain.

Consider performing the $\su{4} \to \su{3} \times \uu{1}$ gauge restriction on the
symmetric product, e.g. $u_2^{(i_1 i_2} (\cO^{(4)}_{36})^{i_3 \cdots i_8)}$,
and extracting terms with power $(\tilde\phi)^5$ in the scalar $\uu{1}$ components.
The graviton piece $u_2^{(i_1 i_2}$ may contribute either zero or two $\uu{1}$
scalar factor $\tilde\phi$, so $\cO^{(4)}_{36}$ must contribute either $5$ or $3$ factors.
Let us write the relevant terms as
\begin{align}\label{u2u3u4decompose}
\begin{split}
 \tr{[\phi^{(i}\phi^{j)}]}
 ~\xrightarrow{~\su{3} \times \uu{1}~}~&
 \tr{[\phi^{(i}\phi^{j)}]}_{\su{3}} + 12\, \tilde\phi^{i} \tilde\phi^{j}~, \\
 \tr{[\phi^{(i}\phi^j\phi^{k)}]}
 ~\xrightarrow{~\su{3} \times \uu{1}~}~&
 \tr{[\phi^{(i}\phi^j\phi^{k)}]}_{\su{3}}
 + 3\, \tilde\phi^{(i} \tr{[\phi^j\phi^{k)}]}_{\su{3}}
 - 24\, \tilde\phi^{i} \tilde\phi^{j} \tilde\phi^{k}~, \\
 \tr{[\phi^{(i}\phi^j\phi^k\phi^{l)}]}
 ~\xrightarrow{~\su{3} \times \uu{1}~}~&
 \tr{[\phi^{(i}\phi^j\phi^k\phi^{l)}]}_{\su{3}}
 + 4\, \tilde\phi^{(i} \tr{[\phi^j\phi^k\phi^{l)}]}_{\su{3}} \\
 & + 6\, \tilde\phi^{(i} \tilde\phi^{j} \tr{[\phi^k\phi^{l)}]}_{\su{3}}
 + 84\, \tilde\phi^{i} \tilde\phi^{j} \tilde\phi^{k} \tilde\phi^{l}~,
\end{split}
\end{align}
and
\be\label{more-SU4-prove}
 (\cO_{36}^{(4)})^{i_1 \cdots i_6} ~\xrightarrow{~\su{3} \times \uu{1}~}~
 \tilde\phi^{a_1} \tilde\phi^{a_2} \tilde\phi^{a_3} \, \tilde\cT^{(i_1 \cdots i_6)}_{(a_1 a_2 a_3)}
 + \tilde\phi^{a_1} \cdots \tilde\phi^{a_5} \, \tilde{\mathcal{S}}^{(i_1 \cdots i_6)}_{(a_1 \cdots a_5)}
 + \text{(other powers of $\tilde\phi$)}~,
\ee
analogous to \eqref{more-B1-prove}.
The $\su{3}$ operator $\tilde{\mathcal{S}}^{(i_1 \cdots i_6)}_{(a_1 \cdots a_5)}$
carries total charge $j=26$, and it is known that there is no fortuitous operator there.
So it is at most a graviton cohomology.
$\tilde\cT^{(i_1 \cdots i_6)}_{(a_1 a_2 a_3)}$ is the tensor that contains
the $\su{3}$ threshold fortuitous cohomology $\cO^{(3)}_{30}$ in its $[3,0]$ part,
as we have just discussed.
The $(\tilde\phi)^5$ term from the gauge restriction of the product
$u_2^{(i_1 i_2} (\cO^{(4)}_{36})^{i_3 \cdots i_8)}$ is therefore
\be
u_2^{(i_1 i_2} (\cO^{(4)}_{36})^{i_3 \cdots i_8)}
~\xrightarrow{~\su{3} \times \uu{1}~}~
12 \, \tilde\phi^{a_1}  \cdots \tilde\phi^{a_3} \tilde\phi^{(i_1} \tilde\phi^{i_2} \, 
\tilde\cT^{i_3 \cdots i_8)}_{(a_1 a_2 a_3)}
+ \tilde\phi^{a_1} \cdots \tilde\phi^{a_5} \, {\rm tr}[\phi^{(i_1}\phi^{i_2}]_{\su{3}}
\tilde{\mathcal{S}}^{i_3 \cdots i_8)}_{(a_1 \cdots a_5)} + \cdots~,
\ee
Extracting the $[3,0]$ part from the $\su{3}$ operator $\tilde\cT$ on the RHS,
we get the same $\cT^{i_1 i_2 i_3}$ from $\tilde\cT$ that we have argued to be
a non-zero multiple of the fortuitous cohomology $\cO^{(3)}_{30}$,
as well as another $[3,0]$ operator involving $\tilde{\mathcal{S}}$ from the second term.
However, we just argued that $\tilde{\mathcal{S}}$ must be a collection of
graviton cohomologies, and the product is also a graviton cohomology.
Thus the sum of two contributions is the fortuitous cohomology $\cO^{(3)}_{30}$,
up to addition of graviton cohomologies.
This proves that the symmetric product $u_2^{(i_1 i_2} (\cO^{(4)}_{36})^{i_3 \cdots i_8)}$
is also a brane-fused operator with the same $\su{3}$ fortuitous cohomology
$\cO^{(3)}_{30}$ in the center, dressed by 5 factors of $\uu{1}$ scalar components.

It is straightforward to generalize the proof to any symmetric products between
$\cO^{(4)}_{36}$ and any number of gravitons $u_2$, $u_3$ and $u_4$.
This accounts for an infinite series of fortuitous operators $[n \geq 8,0]$
with $j = 24+2n$, $Y = 7+n$ and thus $J = j/2-Y = 5$,
in the $\su{4}$ gauge theory.

However, we have reported in Section \ref{sec:SU4census} that there are two copies of
$\su{4}$ fortuitous cohomologies at $[8,0]$ and $j=40$,
whereas the argument above only accounts for one of them.
We find that two independent cohomologies, one of which is the symmetric product
$u_2 \cO^{(4)}_{36}$ discussed above, are both brane-fused black hole operators.

Given two operators at $[8,0]$ and $j=40$ obtained in the course of Section \ref{sec:SU4census},
we perform the $\su{4} \to \su{3} \times \uu{1}$ gauge restriction as usual,
but this time collect \emph{both} $(\tilde\phi)^5$ and $(\tilde\phi)^4$ terms
in the $\uu{1}$ components.
Then we can extract a $[3,0]$ part from the $\su{3}$ operator $[8,0] \otimes [0,5]$,
and a $[4,0]$ part from the $\su{3}$ operator $[8,0] \otimes [0,4]$.
It is possible to choose two independent linear combinations of the two $[8,0]$ operators,
say $\cO^{(4)}_{40a}$ and $\cO^{(4)}_{40b}$, such that the following holds.
\begin{itemize}
\item
$\cO^{(4)}_{40a}$ contains the $\su{3}$ threshold fortuitous cohomology $\cO^{(3)}_{30}$
as the $[3,0]$ part from the $(\tilde\phi)^5$ term.
On the other hand, its $[4,0]$ part from the $(\tilde\phi)^4$ term is $Q$-exact.

\item 
$\cO^{(4)}_{40b}$ contains the $\su{3}$ fortuitous cohomology $\cO^{(3)}_{32}$
as the $[4,0]$ part from the $(\tilde\phi)^4$ term.
On the other hand, its $[3,0]$ part from the $(\tilde\phi)^5$ term is $Q$-exact.
\end{itemize}
The symmetric product operator $u_2 \cO^{(4)}_{36}$ discussed above,
is a linear combination of $\cO^{(4)}_{40a}$ and $\cO^{(4)}_{40b}$.
Therefore, both $\su{4}$ fortuitous cohomology with $[8,0]$ and $j=40$,
are brane-fused black hole operators.

Recall from the end of Section \ref{sec:branefusedintro}
the holographic interpretation of the brane-fused black hole operator
as a microscopic representation of the dual-dressed black hole \cite{Choi:2024xnv}.
A scheme of hairy black holes has been proposed,
where a D3-brane known as the dual giant graviton
\cite{Grisaru:2000zn,Hashimoto:2000zp} wraps a core black hole.
It is well known that the holographic dual of the $\frac12$-BPS dual giant graviton
is the chiral ring operator with totally symmetric Young diagram,
just like the $\uu{1}$ components we extract
in \eqref{decompose30-1} or in \eqref{decompose60}.
Based on this, it is natural to interpret the brane-fused nature of the
$\su{4}$ fortuitous operator as being dual to a core $\su{2}$ black hole
wrapped by two dual giant gravitons,
or a stack of two D3-branes extending along $S^3 \subset {\rm AdS}_5$.
Each D3-brane reduces the flux by 1 inside compared to outside,
thus a stack of two branes make an $\su{2}$ core black hole look like
an $\su{4}$ black hole from outside.

In \cite{deMelloKoch:2024pcs}, a further observation has been made regarding the
size of the core black hole and the $\uu{1}$ part in brane-fused operators.
The apparent fact that at least three symmetric scalars is needed to wrap the
core $\su{2}$ black hole operator, i.e. the absence of the $[1,0]$ at $j=26$
and $[2,0]$ at $j=28$ fortuitous operators in the $F_1$ tower,
signals the finite size of the core black hole in a qualitative sense.
Namely, such fortuitous operators do not exist because the core black hole
cannot be wrapped by a brane that is smaller than itself.
Our observation of the $\su{4}$ double brane-fused black hole operator adds to this claim,
as the `weight' of the second layer of the dual giant graviton is apparently
subject to the same lower bound as the first layer.
It will be very curious to quantify and more rigorously test this idea.

\section*{Acknowledgments}

We thank Sunjin Choi and Seok Kim for discussions.
We thank Jarryd Bath, Sam van Leuven, and Konstantinos Zoubos for sharing their result on non-Coulomb cohomology.
We also thank Nathan Benjamin, Andrea Guerrieri, Victor Rodriguez, David Simmons-Duffin, Yifan Wang, and Zechuan Zheng for sharing their results on BPS cohomology prior to announcement. 
The research of SK, JL, and JS is supported by the National Research Foundation of Korea (NRF) Grants RS-2024-00405629, RS-2026-25482546. 
The research of SL is supported by FWO projects G094523N, G003523N and G0E2723N,
KU Leuven project C16/25/01 and the FWO Junior Postdoctoral Fellowship 1274626N.
The research of JS is also supported by the KAIST-KIAS collaboration program. 
We acknowledge the use of generative AI, especially Claude Code and ChatGPT/Codex, for coding and suggesting efficient algorithms. 

\appendix

\section{Graviton trace relations}\label{app:gravtrrel}
\setlength{\arraycolsep}{2pt}
\renewcommand{\arraystretch}{1}

Of the fundamental graviton trace relations with the $\su{3}$ gauge group,
i.e. the generating set of $Q$-exact graviton operators,
10 may potentially be used in writing down the ansatz of Section \ref{sec:ansatz}
for operators corresponding to the $B_1$ tower, see Table \ref{tab:s-count}.
We list them below. Note that these have already appeared in
\cite{Choi:2023vdm,deMelloKoch:2024pcs} but we slightly lighten the notation
$R_j^{(n_f, n_\psi)} \to R_j^{J}$ where $J = n_\psi + 2n_f$.
{\allowdisplaybreaks
\begin{equation}
\begin{aligned}\label{tr-R}
        &(R_{12}^{1})_{ijk} &=&~ \epsilon_{(i|a_1a_2}\epsilon_{|j|b_1b_2} u^{a_1b_1} {v^{a_2 b_2}}_{|k)} \\
        &(R_{12}^{1})^i_j &=&~  \epsilon_{j a_1 a_2} \left( 4 u^{a_1 b} {v^{i a_2}}_{b} + 3u^{i a_1b} {v^{a_2}}_{b}\right) \\
        &(R_{12}^{1})^{ij}_{kl} &=&~\epsilon_{a_1 a_2 (k}\left(u^{a_1(i}{v^{j)a_2}}_{l)}+
        u^{ij a_1}{v^{a_2}}_{l)}\right) \\
        &(R_{14}^{2})^{i}_{jkl} &=&~\epsilon_{a_1 a_2 (j} {v^{a_1}}_{k} {v^{i a_2}}_{l)}  \\
        &(R_{16}^{2})_i &=&~\epsilon_{i a_1 a_2} \left(12u^{bc}{v^{a_1}}_b{v^{a_2}}_c +13u^{a_1 b}{v^{a_2}}_c{v^c}_b + 12{{v}^{a_1b}}_c{v^{a_2c}}_b \right) \\
        &(R_{16}^{2})^i_{jk} &=&~
        \epsilon_{a_1a_2(j}\!\left(3u^{ib}{v^{a_1}}_{k)}{v^{a_2}}_b \!-\! 
        7u^{ia_1}{v^b}_{k)}{v^{a_2}}_b \!+\! 6u^{a_1b}{v^i}_{k)}{v^{a_2}}_b 
        \!+\! 24{v^{a_1b}}_{k)}{v^{ia_2}}_b \right) \\
        &(R_{16}^{2})^{ij}_{klm} &=&~ \epsilon_{a_1a_2(k}\left({u^{a_1(i}{v^{j)}}_{l}v^{a_2}}_{m)}+ 3{v^{a_1(i}}_{l}{v^{j)a_2}}_{m)}\right) \\
        &(R_{20}^{3})^i &=&~2{v^a}_c{v^b}_a{v^{ic}}_b - 3{v^i}_a{v^c}_b{v^{ab}}_c  \\
        &(R_{22}^{3})^{ij} &=&~u^{ij}{v^a}_b{v^b}_c{v^c}_a -3 u^{a(i}{v^{j)}}_b{v^b}_c{v^c}_a +3 u^{ab}{v^{(i}}_a{v^{j)}}_c{v^c}_b \\
        &(R_{30}^{5})^{ijk} &=&~\epsilon^{a_1a_2(i}{v^{j}}_{a_1}{v^{k)}}_{a_2} {v^{b}}_c{v^{c}}_d{v^{d}}_b\ .
\end{aligned}
\end{equation}
}
From the expressions on the RHS as well as for the equations below,
traces must be subtracted to give tensors with correct irreps,
see (A.3) of \cite{Choi:2023vdm}.
Of these, 6 were used in the selected syzygies \eqref{syz31}-\eqref{syz41}
in Section \ref{sec:B1tower}.
We record the $Q$-pre-images of them as (note that a pre-image of $R^{J}_j$
is $r^{J + 1}_j$, since $Q$ decreases $J$ by 1)
{\allowdisplaybreaks
\begin{equation}
\begin{aligned}\label{tr-r}
        &(r_{12}^{2})_{ijk} \!\! &=&~ \frac{1}{2} \epsilon_{a_1 a_2 (i} \tr \left( \phi^{a_1} \psi_{j}\phi^{a_2} \psi_{k)} \right) \ , \\
        &(r_{12}^{2})^{i}_{j} &=&~  6 \tr \left( \phi^{(i} \phi^{a)} \psi_{(a} \psi_{j)}\right) - 5 \tr\left(\phi^{[i} \psi_{a}\phi^{a]} \psi_{j}\right)\ , \\
        &(r_{12}^{2})^{ij}_{kl} &=&~   \tr \left(\phi^{(i}\phi^{j)} \psi_{(k}\psi_{l)}\right)\ , \\
        &(r_{14}^{3})^i_{jkl} \!\!&=&~  -\frac{1}{2} \tr \left(\phi^i \psi_{(j}\psi_{k}\psi_{l)} \right)\ , \\
        &(r_{16}^{3})^{ij}_{klm} &=&~ -\tr(\phi^{(i}\phi^{j)}\psi_{(k}\psi_{l}\psi_{m)}) -2\tr(\phi^{(i}\psi_{(k}\phi^{j)}\psi_{l}\psi_{m)}) \\
        &(r_{20}^{4})^i &=&~ \epsilon^{a_2 a_3 a_4} \Big[
        \tr(\phi^{[i} \psi_{a_1} \phi^{a_1]} \psi_{a_2} \psi_{a_3} \psi_{a_4})
        - \frac{5}{2} \tr(\phi^{(i} \psi_{a_2} \psi_{a_1} \phi^{a_1)} \psi_{a_3} \psi_{a_4}) \Big] \\
        &&& + \epsilon^{i a_3 a_4} \Big[
        - \frac{17}{8} \tr(\phi^{a_1} \phi^{a_2} \psi_{a_2} \psi_{a_3} \psi_{a_1} \psi_{a_4})
        - \frac{17}{8} \tr(\phi^{a_1} \phi^{a_2} \psi_{a_3} \psi_{a_2} \psi_{a_4} \psi_{a_1})  \\
        &&& + \frac{5}{2} \tr(\phi^{a_1} \phi^{a_2} \psi_{a_2} \psi_{a_3} \psi_{a_4} \psi_{a_1})
        + \tr(\phi^{a_1} \phi^{a_2} \psi_{a_3} \psi_{a_1} \psi_{a_2} \psi_{a_4})
        + \frac{15}{8} \tr(\phi^{a_1} \phi^{a_2} \psi_{a_3} \psi_{a_2} \psi_{a_1} \psi_{a_4})  \\
        &&& + \frac{3}{8} \tr(\phi^{a_1} \psi_{a_1} \psi_{a_2} \psi_{a_3} \phi^{a_2} \psi_{a_4})
        + \frac{3}{8} \tr(\phi^{a_1} \psi_{a_3} \psi_{a_1} \psi_{a_2} \phi^{a_2} \psi_{a_4})
        + \frac{3}{8} \tr(\phi^{a_1} \psi_{a_2} \psi_{a_3} \psi_{a_1} \phi^{a_2} \psi_{a_4}) \\
        &&& - \frac{7}{16} \tr(\phi^{a_1} \phi^{a_2} \psi_{a_1} \psi_{a_2}) \tr(\psi_{a_3} \psi_{a_4})
        - 2 \tr(\phi^{a_1} \phi^{a_2} \psi_{a_3} \psi_{a_4}) \tr(\psi_{a_1} \psi_{a_2}) \\
        &&& - \tr(\phi^{a_1} \phi^{a_2} \psi_{a_3}) \tr(\psi_{a_1} \psi_{a_2} \psi_{a_4})
        + \frac{1}{4} \tr(\phi^{a_1} \phi^{a_2} \psi_{a_3}) \tr(\psi_{a_1} \psi_{a_4} \psi_{a_2}) \\
        &&& + \frac{5}{8} \tr(\phi^{a_1} \psi_{a_1} \psi_{a_3}) \tr(\phi^{a_2} \psi_{a_4} \psi_{a_2})
        + \frac{15}{8} \tr(\phi^{a_1} \psi_{a_2} \psi_{a_3}) \tr(\phi^{a_2} \psi_{a_4} \psi_{a_1}) \Big] \\
        &&& + \frac{5}{4} \tr(\phi^i [\phi^{a_1}, \psi_{a_1}] [\phi^{a_2}, \psi_{a_2}] f)~.
\end{aligned}
\end{equation}
}

Then, non-fundamental graviton trace relations that are products between graviton polynomials
and one of \eqref{tr-R}, are used to write the brane-fused operators.
Consistently with the fact that we only listed fundamental relations $R_i$ that
only involve $u$- and $v$-type gravitons, we only multiply such gravitons to $R_i$.
To describe effectively these products that we call $s_j$,
we present contraction structures in matrix form for each combination of graviton species
and a fundamental graviton trace relation $R$ displayed in Table \ref{tab:s-count}.
For example, the gravitons $(v_2)^2 \, v_3$ times the trace relation $R_{14}^{2}$ with representation $[1,3]$ can yield 24 independent operators with representation $[3,1]$ as desired for constructing the black hole operator $\cO^{(3)}_{34}[3,1]$. They can be described by how 5 pairs of upper and lower indices are contracted inside
\be
\ep^{a_1 a_2 a_3} {(v_2)^{a_4}}_{b_1} {(v_2)^{a_5}}_{b_2} {(v_3)^{a_6 a_7}}_{b_3} {(R_{14}^2)^{a_8}}_{b_4 b_5 b_6}~.
\ee
We express the contraction as, for example,
{\small $\left(
\begin{array}{ccccc}
 1 & 7 & 3 & 5 & 6 \\
 1 & 2 & 4 & 5 & 6 \\
\end{array}
\right)$},
where it means that an upper index $a_i$ is contracted with a lower index $b_j$ where $(i,j)$ is a column of the matrix.
So this example leads to
\be
\ep^{c_1 a_2 c_3} {(v_2)^{a_4}}_{c_1} {(v_2)^{c_4}}_{c_2} {(v_3)^{c_5 c_2}}_{b_3} {(R_{14}^2)^{a_8}}_{c_3 c_4 c_5}~,
\ee
which has 3 free upper indices and 1 free lower index. The free indices must be symmetrized with themselves, and traces must be subtracted, then this would be an operator with the desired order and the desired $[3,1]$ representation. This example gives $s^{[3,1]}_{12}$.

With the notation understood, now we present all the $s$ operators that are used in this paper.
Note that the subscript labels occasionally jump, because we are not displaying the trace relations
that are not used in the construction of any operators in this paper.
The labels are nevertheless sequential according to Table \ref{tab:s-count}.

\paragraph{Using $R_{12}^1[0,3]$}

This fundamental trace relation leads to 2 $s^{[4,1]}$ in the form
\be
\ep^{a_1 a_2 a_3} \ep^{a_4 a_5 a_6} {(v_2)^{a_7}}_{b_1} {(v_2)^{a_8}}_{b_2} {(v_2)^{a_9}}_{b_3} {(v_2)^{a_{10}}}_{b_4} {(R_{12}^1)}_{b_5 b_6 b_7}~,
\ee
with contractions
{\small
\begin{align*}
\left(
\begin{array}{cccccc}
 2 & 3 & 5 & 6 & 7 & 8 \\
 1 & 2 & 3 & 4 & 5 & 6 \\
\end{array}
\right)
&\to s^{[4,1]}_{1}~, &
\left(
\begin{array}{cccccc}
 2 & 4 & 5 & 6 & 3 & 7 \\
 1 & 2 & 3 & 4 & 5 & 6 \\
\end{array}
\right)
&\to s^{[4,1]}_{2}~.
\end{align*}}

\paragraph{Using $R_{12}^1[1,1]$}

This fundamental trace relation leads to 2 $s^{[4,1]}$ in the form
\be
\ep^{a_1 a_2 a_3} {(v_2)^{a_4}}_{b_1} {(v_2)^{a_5}}_{b_2} {(v_2)^{a_6}}_{b_3} {(v_2)^{a_7}}_{b_4} {(R_{12}^1)^{a_8}}_{b_5}~,
\ee
with contractions
{\small
\begin{align*}
\left(
\begin{array}{cccc}
 1 & 2 & 3 & 4 \\
 2 & 3 & 4 & 5 \\
\end{array}
\right)
&\to s^{[4,1]}_{3}~, &
\left(
\begin{array}{cccc}
 2 & 4 & 5 & 3 \\
 2 & 3 & 4 & 5 \\
\end{array}
\right)
&\to s^{[4,1]}_{4}~.
\end{align*}}

\paragraph{Using $R_{12}^1[2,2]$}

This fundamental trace relation leads to 6 $s^{[4,1]}$ in the form
\be
\ep^{a_1 a_2 a_3} {(v_2)^{a_4}}_{b_1} {(v_2)^{a_5}}_{b_2} {(v_2)^{a_6}}_{b_3} {(v_2)^{a_7}}_{b_4} {(R_{12}^1)^{a_8 a_9}}_{b_5 b_6}~,
\ee
with contractions
{\small
\begin{align*}
\left(
\begin{array}{ccccc}
 1 & 2 & 3 & 4 & 7 \\
 1 & 2 & 3 & 4 & 5 \\
\end{array}
\right)
&\to s^{[4,1]}_{5}~, &
\left(
\begin{array}{ccccc}
 1 & 2 & 3 & 8 & 7 \\
 1 & 2 & 3 & 4 & 5 \\
\end{array}
\right)
&\to s^{[4,1]}_{6}~, &
\left(
\begin{array}{ccccc}
 2 & 3 & 4 & 5 & 7 \\
 1 & 2 & 3 & 4 & 5 \\
\end{array}
\right)
&\to s^{[4,1]}_{7}~, \\
\left(
\begin{array}{ccccc}
 3 & 4 & 8 & 2 & 7 \\
 1 & 2 & 3 & 4 & 5 \\
\end{array}
\right)
&\to s^{[4,1]}_{8}~, &
\left(
\begin{array}{ccccc}
 1 & 2 & 3 & 6 & 7 \\
 2 & 3 & 4 & 5 & 6 \\
\end{array}
\right)
&\to s^{[4,1]}_{9}~, &
\left(
\begin{array}{ccccc}
 4 & 1 & 2 & 6 & 7 \\
 2 & 3 & 4 & 5 & 6 \\
\end{array}
\right)
&\to s^{[4,1]}_{10}~.
\end{align*}}

\paragraph{Using $R_{14}^2[1,3]$}

This fundamental trace relation leads to 5 $s^{[2,1]}$ in the form
\be
\ep^{a_1 a_2 a_3} {(v_2)^{a_4}}_{b_1} {(v_2)^{a_5}}_{b_2} {(v_2)^{a_6}}_{b_3} {(R_{14}^2)^{a_7}}_{b_4 b_5 b_6}~,
\ee
with contractions
{\small
\begin{align*}
\left(
\begin{array}{ccccc}
 1 & 2 & 3 & 4 & 5 \\
 1 & 2 & 3 & 4 & 5 \\
\end{array}
\right)
&\to s^{[2,1]}_{1}~, &
\left(
\begin{array}{ccccc}
 1 & 2 & 4 & 3 & 5 \\
 1 & 2 & 3 & 4 & 5 \\
\end{array}
\right)
&\to s^{[2,1]}_{2}~, &
\left(
\begin{array}{ccccc}
 1 & 2 & 3 & 4 & 5 \\
 1 & 2 & 4 & 5 & 6 \\
\end{array}
\right)
&\to s^{[2,1]}_{3}~, \\
\left(
\begin{array}{ccccc}
 1 & 2 & 3 & 4 & 7 \\
 1 & 2 & 4 & 5 & 3 \\
\end{array}
\right)
&\to s^{[2,1]}_{4}~, &
\left(
\begin{array}{ccccc}
 7 & 2 & 3 & 5 & 6 \\
 1 & 2 & 4 & 5 & 6 \\
\end{array}
\right)
&\to s^{[2,1]}_{5}~,
\end{align*}}
24 $s^{[3,1]}$ in the form
\be
\ep^{a_1 a_2 a_3} {(v_2)^{a_4}}_{b_1} {(v_2)^{a_5}}_{b_2} {(v_3)^{a_6 a_7}}_{b_3} {(R_{14}^2)^{a_8}}_{b_4 b_5 b_6}~,
\ee
with contractions
{\small \allowdisplaybreaks
\begin{align*}
\left(
\begin{array}{ccccc}
 1 & 2 & 3 & 6 & 7 \\
 1 & 2 & 3 & 4 & 5 \\
\end{array}
\right)
&\to s^{[3,1]}_{1}~, &
\left(
\begin{array}{ccccc}
 1 & 2 & 8 & 6 & 7 \\
 1 & 2 & 3 & 4 & 5 \\
\end{array}
\right)
&\to s^{[3,1]}_{2}~, &
\left(
\begin{array}{ccccc}
 1 & 2 & 4 & 6 & 7 \\
 1 & 2 & 3 & 4 & 5 \\
\end{array}
\right)
&\to s^{[3,1]}_{3}~, &
\left(
\begin{array}{ccccc}
 1 & 2 & 3 & 4 & 5 \\
 1 & 2 & 3 & 4 & 5 \\
\end{array}
\right)
&\to s^{[3,1]}_{4}~, \\
\left(
\begin{array}{ccccc}
 1 & 2 & 4 & 3 & 5 \\
 1 & 2 & 3 & 4 & 5 \\
\end{array}
\right)
&\to s^{[3,1]}_{5}~, &
\left(
\begin{array}{ccccc}
 1 & 2 & 4 & 3 & 6 \\
 1 & 2 & 3 & 4 & 5 \\
\end{array}
\right)
&\to s^{[3,1]}_{6}~, &
\left(
\begin{array}{ccccc}
 1 & 2 & 4 & 5 & 7 \\
 1 & 2 & 3 & 4 & 5 \\
\end{array}
\right)
&\to s^{[3,1]}_{7}~, &
\left(
\begin{array}{ccccc}
 1 & 2 & 5 & 6 & 7 \\
 1 & 2 & 4 & 5 & 6 \\
\end{array}
\right)
&\to s^{[3,1]}_{8}~, \\
\left(
\begin{array}{ccccc}
 1 & 2 & 3 & 6 & 7 \\
 1 & 2 & 4 & 5 & 6 \\
\end{array}
\right)
&\to s^{[3,1]}_{9}~, &
\left(
\begin{array}{ccccc}
 1 & 2 & 4 & 5 & 6 \\
 1 & 2 & 4 & 5 & 6 \\
\end{array}
\right)
&\to s^{[3,1]}_{10}~, &
\left(
\begin{array}{ccccc}
 1 & 4 & 3 & 5 & 6 \\
 1 & 2 & 4 & 5 & 6 \\
\end{array}
\right)
&\to s^{[3,1]}_{11}~, &
\left(
\begin{array}{ccccc}
 1 & 7 & 3 & 5 & 6 \\
 1 & 2 & 4 & 5 & 6 \\
\end{array}
\right)
&\to s^{[3,1]}_{12}~, \\
\left(
\begin{array}{ccccc}
 1 & 8 & 3 & 5 & 6 \\
 1 & 2 & 4 & 5 & 6 \\
\end{array}
\right)
&\to s^{[3,1]}_{13}~, &
\left(
\begin{array}{ccccc}
 1 & 2 & 3 & 4 & 5 \\
 1 & 2 & 4 & 5 & 6 \\
\end{array}
\right)
&\to s^{[3,1]}_{14}~, &
\left(
\begin{array}{ccccc}
 1 & 2 & 3 & 4 & 6 \\
 1 & 2 & 3 & 4 & 5 \\
\end{array}
\right)
&\to s^{[3,1]}_{15}~, &
\left(
\begin{array}{ccccc}
 1 & 7 & 2 & 4 & 6 \\
 1 & 2 & 3 & 4 & 5 \\
\end{array}
\right)
&\to s^{[3,1]}_{16}~, \\
\left(
\begin{array}{ccccc}
 1 & 2 & 8 & 4 & 6 \\
 1 & 2 & 3 & 4 & 5 \\
\end{array}
\right)
&\to s^{[3,1]}_{17}~, &
\left(
\begin{array}{ccccc}
 1 & 2 & 8 & 3 & 5 \\
 1 & 2 & 3 & 4 & 5 \\
\end{array}
\right)
&\to s^{[3,1]}_{18}~, &
\left(
\begin{array}{ccccc}
 1 & 6 & 2 & 3 & 5 \\
 1 & 2 & 3 & 4 & 5 \\
\end{array}
\right)
&\to s^{[3,1]}_{19}~, &
\left(
\begin{array}{ccccc}
 1 & 8 & 2 & 3 & 5 \\
 1 & 2 & 3 & 4 & 5 \\
\end{array}
\right)
&\to s^{[3,1]}_{20}~, \\
\left(
\begin{array}{ccccc}
 1 & 4 & 8 & 3 & 5 \\
 1 & 2 & 3 & 4 & 5 \\
\end{array}
\right)
&\to s^{[3,1]}_{21}~, &
\left(
\begin{array}{ccccc}
 1 & 6 & 8 & 3 & 5 \\
 1 & 2 & 3 & 4 & 5 \\
\end{array}
\right)
&\to s^{[3,1]}_{22}~, &
\left(
\begin{array}{ccccc}
 1 & 2 & 8 & 3 & 7 \\
 1 & 2 & 3 & 4 & 5 \\
\end{array}
\right)
&\to s^{[3,1]}_{23}~, &
\left(
\begin{array}{ccccc}
 1 & 2 & 3 & 4 & 6 \\
 2 & 3 & 4 & 5 & 6 \\
\end{array}
\right)
&\to s^{[3,1]}_{24}~,
\end{align*}
}
and 32 $s^{[4,1]}$ in the form
\be
\ep^{a_1 a_2 a_3} {(v_2)^{a_4}}_{b_1} {(v_3)^{a_5 a_6}}_{b_2} {(v_3)^{a_7 a_8}}_{b_3} {(R_{14}^2)^{a_9}}_{b_4 b_5 b_6}~,
\ee
with contractions
{\small \allowdisplaybreaks
\begin{align*}
\left(
\begin{array}{ccccc}
 1 & 2 & 3 & 7 & 8 \\
 1 & 2 & 3 & 4 & 5 \\
\end{array}
\right)
&\to s^{[4,1]}_{26}~, &
\left(
\begin{array}{ccccc}
 1 & 2 & 4 & 7 & 8 \\
 1 & 2 & 3 & 4 & 5 \\
\end{array}
\right)
&\to s^{[4,1]}_{27}~, &
\left(
\begin{array}{ccccc}
 1 & 2 & 5 & 7 & 8 \\
 1 & 2 & 3 & 4 & 5 \\
\end{array}
\right)
&\to s^{[4,1]}_{28}~, &
\left(
\begin{array}{ccccc}
 1 & 2 & 9 & 7 & 8 \\
 1 & 2 & 3 & 4 & 5 \\
\end{array}
\right)
&\to s^{[4,1]}_{29}~, \\
\left(
\begin{array}{ccccc}
 1 & 9 & 2 & 7 & 8 \\
 1 & 2 & 3 & 4 & 5 \\
\end{array}
\right)
&\to s^{[4,1]}_{30}~, &
\left(
\begin{array}{ccccc}
 1 & 2 & 3 & 6 & 8 \\
 1 & 2 & 3 & 4 & 5 \\
\end{array}
\right)
&\to s^{[4,1]}_{31}~, &
\left(
\begin{array}{ccccc}
 1 & 2 & 5 & 6 & 8 \\
 1 & 2 & 3 & 4 & 5 \\
\end{array}
\right)
&\to s^{[4,1]}_{32}~, &
\left(
\begin{array}{ccccc}
 1 & 2 & 9 & 6 & 8 \\
 1 & 2 & 3 & 4 & 5 \\
\end{array}
\right)
&\to s^{[4,1]}_{33}~, \\
\left(
\begin{array}{ccccc}
 1 & 2 & 3 & 4 & 8 \\
 1 & 2 & 3 & 4 & 5 \\
\end{array}
\right)
&\to s^{[4,1]}_{34}~, &
\left(
\begin{array}{ccccc}
 1 & 2 & 5 & 4 & 8 \\
 1 & 2 & 3 & 4 & 5 \\
\end{array}
\right)
&\to s^{[4,1]}_{35}~, &
\left(
\begin{array}{ccccc}
 1 & 7 & 2 & 4 & 8 \\
 1 & 2 & 3 & 4 & 5 \\
\end{array}
\right)
&\to s^{[4,1]}_{36}~, &
\left(
\begin{array}{ccccc}
 1 & 2 & 9 & 4 & 8 \\
 1 & 2 & 3 & 4 & 5 \\
\end{array}
\right)
&\to s^{[4,1]}_{37}~, \\
\left(
\begin{array}{ccccc}
 1 & 9 & 2 & 4 & 8 \\
 1 & 2 & 3 & 4 & 5 \\
\end{array}
\right)
&\to s^{[4,1]}_{38}~, &
\left(
\begin{array}{ccccc}
 1 & 2 & 4 & 3 & 8 \\
 1 & 2 & 3 & 4 & 5 \\
\end{array}
\right)
&\to s^{[4,1]}_{39}~, &
\left(
\begin{array}{ccccc}
 1 & 2 & 5 & 3 & 8 \\
 1 & 2 & 3 & 4 & 5 \\
\end{array}
\right)
&\to s^{[4,1]}_{40}~, &
\left(
\begin{array}{ccccc}
 1 & 7 & 2 & 3 & 8 \\
 1 & 2 & 3 & 4 & 5 \\
\end{array}
\right)
&\to s^{[4,1]}_{41}~, \\
\left(
\begin{array}{ccccc}
 1 & 2 & 9 & 3 & 8 \\
 1 & 2 & 3 & 4 & 5 \\
\end{array}
\right)
&\to s^{[4,1]}_{42}~, &
\left(
\begin{array}{ccccc}
 1 & 9 & 2 & 3 & 8 \\
 1 & 2 & 3 & 4 & 5 \\
\end{array}
\right)
&\to s^{[4,1]}_{43}~, &
\left(
\begin{array}{ccccc}
 1 & 2 & 5 & 3 & 4 \\
 1 & 2 & 3 & 4 & 5 \\
\end{array}
\right)
&\to s^{[4,1]}_{44}~, &
\left(
\begin{array}{ccccc}
 1 & 2 & 9 & 3 & 4 \\
 1 & 2 & 3 & 4 & 5 \\
\end{array}
\right)
&\to s^{[4,1]}_{45}~, \\
\left(
\begin{array}{ccccc}
 1 & 2 & 6 & 7 & 8 \\
 1 & 2 & 4 & 5 & 6 \\
\end{array}
\right)
&\to s^{[4,1]}_{46}~, &
\left(
\begin{array}{ccccc}
 1 & 2 & 5 & 6 & 8 \\
 1 & 2 & 4 & 5 & 6 \\
\end{array}
\right)
&\to s^{[4,1]}_{47}~, &
\left(
\begin{array}{ccccc}
 1 & 2 & 4 & 7 & 8 \\
 1 & 2 & 4 & 5 & 6 \\
\end{array}
\right)
&\to s^{[4,1]}_{48}~, &
\left(
\begin{array}{ccccc}
 1 & 2 & 4 & 6 & 8 \\
 1 & 2 & 4 & 5 & 6 \\
\end{array}
\right)
&\to s^{[4,1]}_{49}~, \\
\left(
\begin{array}{ccccc}
 1 & 2 & 4 & 5 & 6 \\
 1 & 2 & 4 & 5 & 6 \\
\end{array}
\right)
&\to s^{[4,1]}_{50}~, &
\left(
\begin{array}{ccccc}
 1 & 2 & 3 & 7 & 8 \\
 1 & 2 & 4 & 5 & 6 \\
\end{array}
\right)
&\to s^{[4,1]}_{51}~, &
\left(
\begin{array}{ccccc}
 1 & 9 & 3 & 7 & 8 \\
 1 & 2 & 4 & 5 & 6 \\
\end{array}
\right)
&\to s^{[4,1]}_{52}~, &
\left(
\begin{array}{ccccc}
 1 & 2 & 3 & 6 & 8 \\
 1 & 2 & 4 & 5 & 6 \\
\end{array}
\right)
&\to s^{[4,1]}_{53}~, \\
\left(
\begin{array}{ccccc}
 1 & 2 & 3 & 4 & 8 \\
 1 & 2 & 4 & 5 & 6 \\
\end{array}
\right)
&\to s^{[4,1]}_{54}~, &
\left(
\begin{array}{ccccc}
 1 & 7 & 3 & 4 & 8 \\
 1 & 2 & 4 & 5 & 6 \\
\end{array}
\right)
&\to s^{[4,1]}_{55}~, &
\left(
\begin{array}{ccccc}
 1 & 9 & 3 & 4 & 8 \\
 1 & 2 & 4 & 5 & 6 \\
\end{array}
\right)
&\to s^{[4,1]}_{56}~, &
\left(
\begin{array}{ccccc}
 1 & 2 & 3 & 5 & 6 \\
 1 & 2 & 4 & 5 & 6 \\
\end{array}
\right)
&\to s^{[4,1]}_{57}~.
\end{align*}
}

\paragraph{Using $R_{16}^2[2,3]$}

This fundamental trace relation leads to 5 $s^{[3,1]}$ in the form
\be
\ep^{a_1 a_2 a_3} {(v_2)^{a_4}}_{b_1} {(v_2)^{a_5}}_{b_2} {(v_2)^{a_6}}_{b_3} {(R_{16}^2)^{a_7 a_8}}_{b_4 b_5 b_6}~,
\ee
with contractions
{\small
\begin{align*}
\left(
\begin{array}{ccccc}
 1 & 2 & 3 & 5 & 6 \\
 2 & 3 & 4 & 5 & 6 \\
\end{array}
\right)
&\to s^{[3,1]}_{32}~, &
\left(
\begin{array}{ccccc}
 1 & 2 & 3 & 4 & 5 \\
 2 & 3 & 4 & 5 & 6 \\
\end{array}
\right)
&\to s^{[3,1]}_{33}~, &
\left(
\begin{array}{ccccc}
 1 & 7 & 3 & 4 & 5 \\
 2 & 3 & 4 & 5 & 6 \\
\end{array}
\right)
&\to s^{[3,1]}_{34}~, \\
\left(
\begin{array}{ccccc}
 1 & 2 & 4 & 3 & 6 \\
 1 & 2 & 3 & 5 & 6 \\
\end{array}
\right)
&\to s^{[3,1]}_{35}~, &
\left(
\begin{array}{ccccc}
 1 & 2 & 7 & 3 & 6 \\
 1 & 2 & 3 & 5 & 6 \\
\end{array}
\right)
&\to s^{[3,1]}_{36}~,
\end{align*}}
and 25 $s^{[4,1]}$ in the form
\be
\ep^{a_1 a_2 a_3} {(v_2)^{a_4}}_{b_1} {(v_2)^{a_5}}_{b_2} {(v_3)^{a_6 a_7}}_{b_3} {(R_{16}^2)^{a_8 a_9}}_{b_4 b_5 b_6}~,
\ee
with contractions
{\small \allowdisplaybreaks
\begin{align*}
\left(
\begin{array}{ccccc}
 1 & 2 & 5 & 6 & 7 \\
 2 & 3 & 4 & 5 & 6 \\
\end{array}
\right)
&\to s^{[4,1]}_{82}~, &
\left(
\begin{array}{ccccc}
 1 & 2 & 5 & 6 & 7 \\
 1 & 2 & 4 & 5 & 6 \\
\end{array}
\right)
&\to s^{[4,1]}_{83}~, &
\left(
\begin{array}{ccccc}
 1 & 2 & 3 & 6 & 7 \\
 2 & 3 & 4 & 5 & 6 \\
\end{array}
\right)
&\to s^{[4,1]}_{84}~, &
\left(
\begin{array}{ccccc}
 1 & 4 & 3 & 6 & 7 \\
 2 & 3 & 4 & 5 & 6 \\
\end{array}
\right)
&\to s^{[4,1]}_{85}~, \\
\left(
\begin{array}{ccccc}
 1 & 8 & 3 & 6 & 7 \\
 2 & 3 & 4 & 5 & 6 \\
\end{array}
\right)
&\to s^{[4,1]}_{86}~, &
\left(
\begin{array}{ccccc}
 1 & 2 & 3 & 6 & 7 \\
 1 & 2 & 4 & 5 & 6 \\
\end{array}
\right)
&\to s^{[4,1]}_{87}~, &
\left(
\begin{array}{ccccc}
 1 & 2 & 4 & 5 & 7 \\
 2 & 3 & 4 & 5 & 6 \\
\end{array}
\right)
&\to s^{[4,1]}_{88}~, &
\left(
\begin{array}{ccccc}
 1 & 2 & 3 & 5 & 7 \\
 2 & 3 & 4 & 5 & 6 \\
\end{array}
\right)
&\to s^{[4,1]}_{89}~, \\
\left(
\begin{array}{ccccc}
 1 & 4 & 3 & 5 & 7 \\
 2 & 3 & 4 & 5 & 6 \\
\end{array}
\right)
&\to s^{[4,1]}_{90}~, &
\left(
\begin{array}{ccccc}
 1 & 8 & 3 & 5 & 7 \\
 2 & 3 & 4 & 5 & 6 \\
\end{array}
\right)
&\to s^{[4,1]}_{91}~, &
\left(
\begin{array}{ccccc}
 1 & 2 & 3 & 5 & 7 \\
 1 & 2 & 4 & 5 & 6 \\
\end{array}
\right)
&\to s^{[4,1]}_{92}~, &
\left(
\begin{array}{ccccc}
 1 & 6 & 3 & 5 & 7 \\
 1 & 2 & 4 & 5 & 6 \\
\end{array}
\right)
&\to s^{[4,1]}_{93}~, \\
\left(
\begin{array}{ccccc}
 1 & 8 & 3 & 5 & 7 \\
 1 & 2 & 4 & 5 & 6 \\
\end{array}
\right)
&\to s^{[4,1]}_{94}~, &
\left(
\begin{array}{ccccc}
 1 & 2 & 3 & 4 & 5 \\
 2 & 3 & 4 & 5 & 6 \\
\end{array}
\right)
&\to s^{[4,1]}_{95}~, &
\left(
\begin{array}{ccccc}
 1 & 8 & 3 & 4 & 5 \\
 2 & 3 & 4 & 5 & 6 \\
\end{array}
\right)
&\to s^{[4,1]}_{96}~, &
\left(
\begin{array}{ccccc}
 1 & 2 & 3 & 4 & 5 \\
 1 & 2 & 4 & 5 & 6 \\
\end{array}
\right)
&\to s^{[4,1]}_{97}~, \\
\left(
\begin{array}{ccccc}
 1 & 2 & 3 & 5 & 7 \\
 1 & 2 & 3 & 5 & 6 \\
\end{array}
\right)
&\to s^{[4,1]}_{98}~, &
\left(
\begin{array}{ccccc}
 1 & 2 & 8 & 5 & 7 \\
 1 & 2 & 3 & 5 & 6 \\
\end{array}
\right)
&\to s^{[4,1]}_{99}~, &
\left(
\begin{array}{ccccc}
 1 & 2 & 4 & 3 & 7 \\
 1 & 2 & 3 & 5 & 6 \\
\end{array}
\right)
&\to s^{[4,1]}_{100}~, \\
\left(
\begin{array}{ccccc}
 1 & 2 & 8 & 3 & 7 \\
 1 & 2 & 3 & 5 & 6 \\
\end{array}
\right)
&\to s^{[4,1]}_{101}~, &
\left(
\begin{array}{ccccc}
 1 & 2 & 4 & 3 & 5 \\
 1 & 2 & 3 & 5 & 6 \\
\end{array}
\right)
&\to s^{[4,1]}_{102}~, &
\left(
\begin{array}{ccccc}
 1 & 2 & 8 & 3 & 5 \\
 1 & 2 & 3 & 5 & 6 \\
\end{array}
\right)
&\to s^{[4,1]}_{103}~, \\
\left(
\begin{array}{ccccc}
 1 & 6 & 8 & 3 & 5 \\
 1 & 2 & 3 & 5 & 6 \\
\end{array}
\right)
&\to s^{[4,1]}_{104}~, &
\left(
\begin{array}{ccccc}
 1 & 6 & 2 & 3 & 5 \\
 1 & 2 & 3 & 5 & 6 \\
\end{array}
\right)
&\to s^{[4,1]}_{105}~, &
\left(
\begin{array}{ccccc}
 1 & 8 & 9 & 3 & 5 \\
 1 & 2 & 3 & 5 & 6 \\
\end{array}
\right)
&\to s^{[4,1]}_{106}~.
\end{align*}
}

\paragraph{Using $R_{20}^3[1,0]$}

This fundamental trace relation leads to
4 $s^{[3,1]}$ in the form
\be
{(v_2)^{a_1}}_{b_1} {(v_3)^{a_2 a_3}}_{b_2} {(R_{20}^3)^{a_4}}~,
\ee
with contractions
\begin{align*}
\left(
\begin{array}{c}
 2 \\
 1 \\
\end{array}
\right)
&\to s^{[3,1]}_{37}~, &
\left(
\begin{array}{c}
 4 \\
 1\\
\end{array}
\right)
&\to s^{[3,1]}_{38}~, &
\left(
\begin{array}{c}
 1 \\
 2 \\
\end{array}
\right)
&\to s^{[3,1]}_{39}~, &
\left(
\begin{array}{c}
 4 \\
 2\\
\end{array}
\right)
&\to s^{[3,1]}_{40}~, &
\end{align*}
3 $s^{[4,1]}$ in the form
\be
(u_2)^{a_1 a_2} {(v_2)^{a_3}}_{b_1} {(v_2)^{a_4}}_{b_2} {(R_{20}^3)^{a_5}}~,
\ee
with contractions
\begin{align*}
\left(
\begin{array}{c}
 1 \\
 1 \\
\end{array}
\right)
&\to s^{[4,1]}_{107}~, &
\left(
\begin{array}{c}
 4 \\
 1\\
\end{array}
\right)
&\to s^{[4,1]}_{108}~, &
\left(
\begin{array}{c}
 5 \\
 1 \\
\end{array}
\right)
&\to s^{[4,1]}_{109}~.
\end{align*}
and 2 more $s^{[4,1]}$ in the form
\be
{(v_3)^{a_1 a_2}}_{b_1} {(v_3)^{a_3 a_4}}_{b_2} {(R_{20}^3)^{a_5}}~,
\ee
with contractions
\begin{align*}
\left(
\begin{array}{c}
 3 \\
 1 \\
\end{array}
\right)
&\to s^{[4,1]}_{110}~, &
\left(
\begin{array}{c}
 5 \\
 1 \\
\end{array}
\right)
&\to s^{[4,1]}_{111}~.
\end{align*}
%

{\footnotesize
\bibliography{Refs}
}

\bibliographystyle{utphys}

\end{document}